\documentclass[english]{article}
\usepackage[T1]{fontenc}
\usepackage[latin9]{inputenc}
\usepackage{geometry}
\usepackage{mathrsfs}
\usepackage{amsmath}
\usepackage{amssymb}
\usepackage{authblk}
\usepackage{graphicx}
\usepackage{esint}
\usepackage{url}
\usepackage{float}
\usepackage{ulem}

\usepackage{cite}

\usepackage{hyperref}
\usepackage{enumitem}

\usepackage[labelfont=bf,labelsep=period,justification=raggedright]{caption}

\makeatletter
\renewcommand{\@biblabel}[1]{\quad#1.}
\makeatother

\date{}

\makeatletter
\renewcommand*{\@fnsymbol}[1]{\ensuremath{\ifcase#1\or *\or \dagger\or \ddagger\or
   \mathsection\or \mathparagraph\or \|\or \forall\or \dagger\dagger
   \or \ddagger\ddagger \else\@ctrerr\fi}}
\makeatother

\usepackage{babel}

\begin{document}

\title{Sparse and compositionally robust inference of microbial ecological networks}

\author[1]{Zachary D. Kurtz\thanks{These authors contributed equally to this work} \thanks{Zachary.Kurtz@med.nyu.edu}}
\author[2,3]{Christian L. Mueller$^*$ \thanks{cm192@nyu.edu}}
\author[1,2,4]{Emily R. Miraldi$^*$ \thanks{emiraldi@nyu.edu}}
\author[1]{Dan R. Littman\thanks{Dan.Littman@med.nyu.edu}}
\author[1]{Martin J. Blaser\thanks{Martin.Blaser@med.nyu.edu}}
\author[2,3,4]{Richard A. Bonneau\thanks{rb133@nyu.edu}}
\affil[1]{\small{Departments of Microbiology and Medicine, New York University School of Medicine, New York, NY 10016}}
\affil[2]{Department of Biology, Center for Genomics and Systems Biology,  New York University, New York, NY 10003}
\affil[3]{Courant Institute of Mathematical Sciences, New York University, New York, NY 10012}
\affil[4]{Simons Center for Data Analysis, Simons Foundation, New York, NY 10010}

\renewcommand\Authands{ and }
\maketitle

\section*{Abstract}
16S ribosomal RNA (rRNA) gene and other environmental sequencing techniques provide snapshots of microbial communities, revealing phylogeny and the abundances of microbial populations across diverse ecosystems. While changes in microbial community structure are demonstrably associated with certain environmental conditions (from metabolic and immunological health in mammals to ecological stability in soils and oceans), identification of underlying mechanisms requires new statistical tools, as these datasets present several technical challenges.  First, the abundances of microbial operational taxonomic units (OTUs) from amplicon-based datasets are compositional. Counts are normalized to the total number of counts in the sample. Thus, microbial abundances are not independent, and traditional statistical metrics (e.g., correlation) for the detection of OTU-OTU relationships can lead to spurious results. Secondly, microbial sequencing-based studies typically measure hundreds of OTUs on only tens to hundreds of samples; thus, inference of OTU-OTU association networks is severely under-powered, and additional information (or assumptions) are required for accurate inference. Here, we present SPIEC-EASI (\textbf{SP}arse \textbf{I}nvers\textbf{E} \textbf{C}ovariance Estimation for \textbf{E}cological \textbf{A}ssociation \textbf{I}nference), a statistical method for the inference of microbial ecological networks from amplicon sequencing datasets that addresses both of these issues. SPIEC-EASI combines data transformations developed for compositional data analysis with a graphical model inference framework that assumes the underlying ecological association network is sparse. To reconstruct the network, SPIEC-EASI relies on algorithms for sparse neighborhood and inverse covariance selection. To provide a synthetic benchmark in the absence of an experimentally validated gold-standard network, SPIEC-EASI is accompanied by a set of computational tools to generate OTU count data from a set of diverse underlying network topologies. SPIEC-EASI outperforms state-of-the-art methods to recover edges and network properties on synthetic data under a variety of scenarios. SPIEC-EASI also reproducibly predicts previously unknown microbial associations using data from the American Gut project.

\section*{Introduction}
\label{sec:intro}

Low-cost metagenomic and amplicon-based sequencing promises to make the resolution of complex interactions between microbial populations and their surrounding environment a routine component of observational ecology and experimental biology. Indeed, large-scale data collection efforts (such as Earth Microbiome Project \cite{Gilbert:2010}, the Human Microbiome Project \cite{Turnbaugh:2007}, and the American Gut Project \cite{AmGutProj}) bring an ever-increasing number of samples from soil, marine and animal-associated microbiota to the public domain. Recent research efforts in ecology, statistics, and computational biology have been aimed at reliably inferring novel biological insights and testable hypotheses from population abundances and phylogenies. Classic objectives in community ecology include, (i) the accurate estimation of the number of taxa (observed and unobserved) from microbial studies \cite{Bunge:2014} and, related to that, (ii) the estimation of community diversity within and across different habitats from the modeled population counts \cite{Foster2008}. Moreover, some microbial compositions appear to form distinct clusters, leading to the concept of enterotypes, or ecological steady states in the gut \cite{Arumugam:2011}, but their existence has not been established with certainty \cite{Koren:2013}. Another aim of recent studies is the elucidation of connections between microbes and environmental or host covariates. Examples include a novel statistical regression framework for relating microbiome compositions and covariates in the context of nutrient intake \cite{Chen:2013}, observations that microbiome compositions strongly correlate with disease status in new-onset Crohn's disease \cite{Gevers:2014}, and the connections between helminth infection and the microbiome diversity \cite{Lee:2014}. 

One goal of microbiome studies is the accurate inference of microbial ecological interactions from population-level data \cite{Faust:2012a}. 'Interactions' are inferred by detecting significant (typically non-directional) associations between sampled populations, e.g., by measuring frequency of co-occurrence \cite{Fuhrman:2008, Barberan:2012}. Microbiota are measured by profiling variable regions of bacterial 16S rRNA gene sequences. These regions are amplified, sequenced, and the resulting reads are then grouped into common Operational Taxonomic Units (OTUs) and quantified, with OTU counts serving as a proxy to the underlying microbial populations' abundances. Knowledge of interaction networks (here, a measure of microbial association) provides a foundation to predictively model the interplay between environment and microbial populations. A recent example is the successful construction of a dynamic differential equation model to describe the primary succession of intestinal microbiota in mice \cite{Marino2014}. A commonly used tool to infer a network is correlation analysis; that is computing Pearson's correlation coefficient among all pairs of OTU samples, and an interaction between microbes is assumed when the absolute value of the correlation coefficient is sufficiently high \cite{Deng:2012,Steele2011}.

However, applying traditional correlation analysis to amplicon surveys of microbial population data is likely to yield spurious results \cite{Friedman2012,Gevers:2014}. To limit experimental biases due to sampling depth, OTU count data is typically transformed by normalizing each OTU count to the total sum of counts in the sample. Thus, communities of microbial relative abundances, termed compositions, are not independent, and classical correlation analysis may fail \cite{Aitchison1981}. Recent methods such as Sparse Correlations for Compositional data (SparCC) \cite{Friedman2012} and Compositionally Corrected by REnormalization and PErmutation (CCREPE) \cite{Faust:2012a, Gevers:2014, Faust:2012b} are designed to account for these compositional biases and represent the state of the art in the field. Yet, it is not clear that correlation is the proper measure of association. For example, correlations can arise between OTUs that are indirectly connected in an ecological network (we expand on this point below).

Dimensionality poses another challenge to statistical analysis of microbiome studies, as the number of measured OTUs $p$ is on the order of  hundreds to thousands whereas the number of samples $n$ generally ranges from tens to hundreds. This implies that any meaningful interaction inference scheme must operate in the underdetermined data regime ($p>n$), which is viable only if additional assumptions about the interaction network can be made. As technological developments lead to greater sequencing depths, new computational methods that address the ($p>n$) challenge will become increasingly important.
\begin{figure}
   \includegraphics[scale=.4]{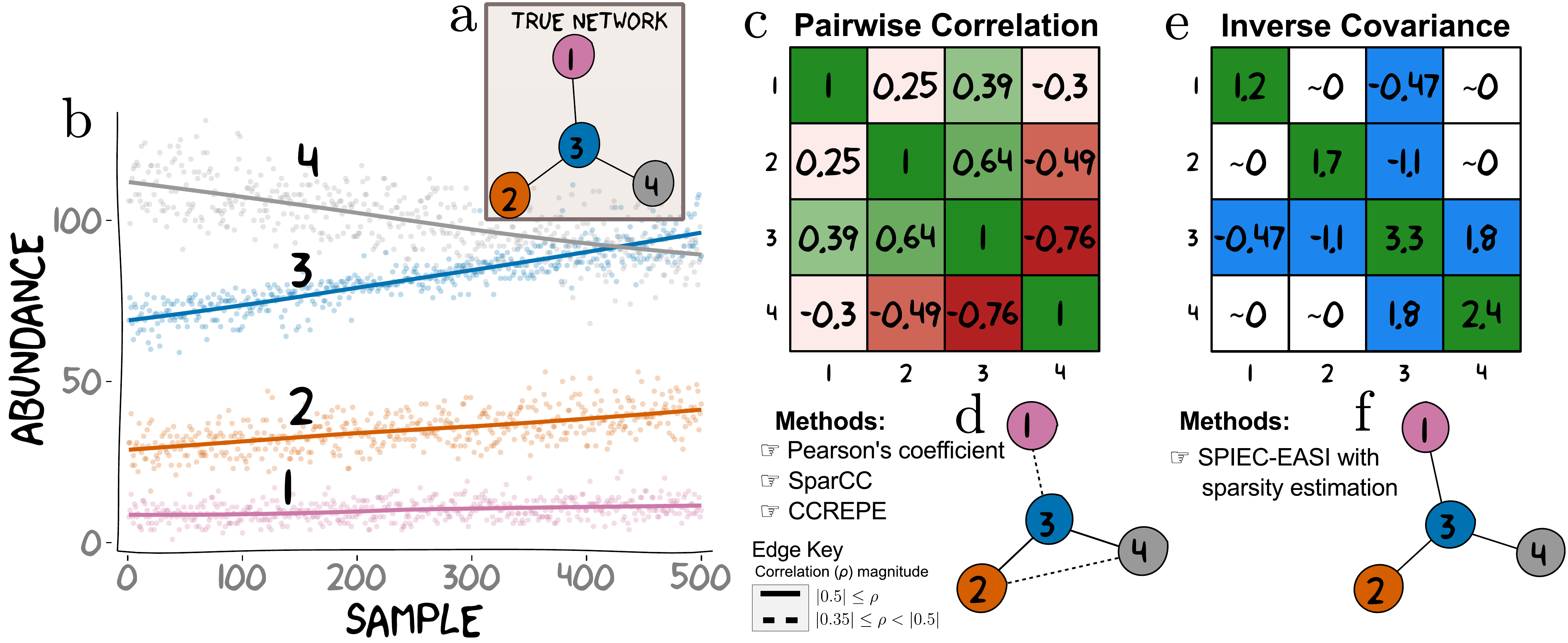}
	\caption{\textbf{Conditional independence vs correlation analysis for a toy dataset:} In an ecosystem, the abundance of any OTU is potentially dependent on the abundances of other OTUs in the ecological network. Here, we simulate abundances from a network where OTU 3 directly influences (via some set of biological mechanisms) the abundances of OTUs 1, 2 and 4 (\textbf{a}). The inference goal here is to recover the underlying network from the simulated data. \textbf{b}) Absolute abundances of these four OTUs were drawn from a negative-binomial distribution across 500 samples according to the true network (as described in the Methods section). \textbf{c}) Computing all pairwise Pearson correlation yields a symmetric matrix showing patterns of association (positive correlations are green and negative are red). We thresholded entries of the correlation matrix to generate relevance networks. \textbf{d}) A threshold at  $\rho \geq |0.35|$ (represented by dashed and solid edges) results in a network in which OTU 3 is connected to all other OTUs with an additional connection between OTU 2 and OTU 4. A more stringent threshold at $\rho \geq |0.5|$, results in a sparser relevance network (notably missing the edge between OTU 3 and OTU 1), and is represented in \textbf{d} by solid edges only. Importantly, no single threshold recovers the true underlying hub topology. \textbf{e}) The inverse sample covariance matrix yields a symmetric matrix where entries are approximately zero if the corresponding OTU pairs are conditionally independent. The network (\textbf{f}) inferred from the non-zero entries (colored in blue in \textbf{e}) identifies the correct hub network. Thus, it is possible to choose a threshold for the sample inverse covariance that faithfully recovers the true network. Such a threshold is not guaranteed to exist for correlation or covariance (the metric used by SparCC and CCREPE). Intuitively, this is because simultaneous direct connections can induce strong correlations between nodes that do not have direct relationships (e.g. OTU 2-4). Conversely, weak correlations can arise between directly connected nodes (e.g. OTU 1-3). Although correlation is a useful measure of association in many contexts, it is a pairwise metric and therefore limited in a multivariate setting. On the other hand, SPIEC-EASI's estimate of entries in the inverse covariance matrix depend on the conditional states of all available nodes. This feature helps SPIEC-EASI avoid detection of indirect network interactions.}
\label{toydata_all}
\end{figure}

In the present work, we present a novel strategy to infer networks from (potentially high-dimensional) community composition data. We introduce SPIEC-EASI (\textbf{SP}arse \textbf{I}nvers\textbf{E} \textbf{C}ovariance Estimation for \textbf{E}cological \textbf{AS}sociation \textbf{I}nference, pronounced \emph{speakeasy}), a new statistical method for the inference of microbial ecological networks and generation of realistic synthetic data. SPIEC-EASI inference comprises two steps: First, a transformation from the field of compositional data analysis is applied to the OTU data. Second, SPIEC-EASI estimates the interaction graph from the transformed data using one of two methods: (i) neighborhood selection \cite{Meinshausen2006,Bonneau:2006} and (ii) sparse inverse covariance selection \cite{Friedman2008,Banerjee2008}. Unlike empirical correlation or covariance estimation, used in SparCC and CCREPE, our pipeline seeks to infer an underlying graphical model using the concept of conditional independence. 

Informally, two nodes (e.g. OTUs) are conditionally independent if, given the state (e.g. abundance) of all other nodes in the network, neither node provides additional information about the state of the other. A link between any two nodes in the graphical model implies that the OTU abundances are not conditionally independent and that there is a (linear) relationship between them that cannot be better explained by an alternate network wiring. In this way, our method avoids detection of correlated but indirectly connected OTUs, thus ensuring parsimony of the resulting network model (for more detail, see \mbox{\nameref{sec:methods}} and Figure \ref{toydata_all}). This model is an undirected graph where links between nodes represent signed associations between OTUs. The use of graphical models has gained considerable popularity in network biology \cite{Wille:2004, Friedman:2004,Bonneau2008} and, more recently, in structural biology \cite{Jones:2011}, particularly to correct for transitive correlations in protein structure prediction \cite{Marks:2012}.

%

To properly benchmark our inference scheme and compare its performance with other state-of-the-art schemes \cite{Friedman2012,Gevers:2014}, SPIEC-EASI is accompanied by a synthetic data generation routine, which generates realistic synthetic OTU data from networks with diverse topologies. This is significant because, to date, (i) no experimentally verified set of ``gold-standard" microbial interactions exists, (ii) previous synthetic benchmark data do not accurately reflect the actual properties of microbiome data \cite{Friedman2012}, and (iii) theoretical and empirical work from high-dimensional statistics \cite{Ravikumar2011,Tandon2014,Liu:2011} suggests that network topology can strongly impact network recovery and performance and thus must be considered in the design of synthetic datasets.
 
We show that SPIEC-EASI is a scalable inference engine that (i) yields superior performance with respect to state-of-the-art methods in terms of interaction recovery and network features in a diverse set of realistic synthetic benchmark scenarios, (ii) provides the most stable and reproducible network when applied to real data, and (iii) reliably estimates an invertible covariance matrix which can be used for additional downstream statistical analysis. In agreement with statistical theory \cite{Ravikumar2011}, inference on the synthetic datasets demonstrates that the degree distribution of the underlying network has the largest effect on performance, and this effect is observed across all methods tested. SPIEC-EASI network inference applied to actual data from the American Gut Project (AGP) shows (i) that clusters of strongly connected components are likely to contain OTUs with common family membership and (ii) that actual gut microbial networks are likely composites of archetypical network topologies. 
In the \nameref{sec:methods} section, we present statistical and computational aspects of SPIEC-EASI. We then benchmark SPIEC-EASI, comparing it to current inference schemes using synthetic data. We then apply SPIEC-EASI to measurements available from the AGP database. The SPIEC-EASI pipeline is implemented in the R package [SpiecEasi] freely available at \url{https://github.com/zdk123/SpiecEasi}. All presented numerical data is available at \url{http://bonneaulab.bio.nyu.edu/}.

\section*{Materials and Methods}
\label{sec:methods}

\begin{figure}[H]
\centering
\includegraphics[scale=0.4]{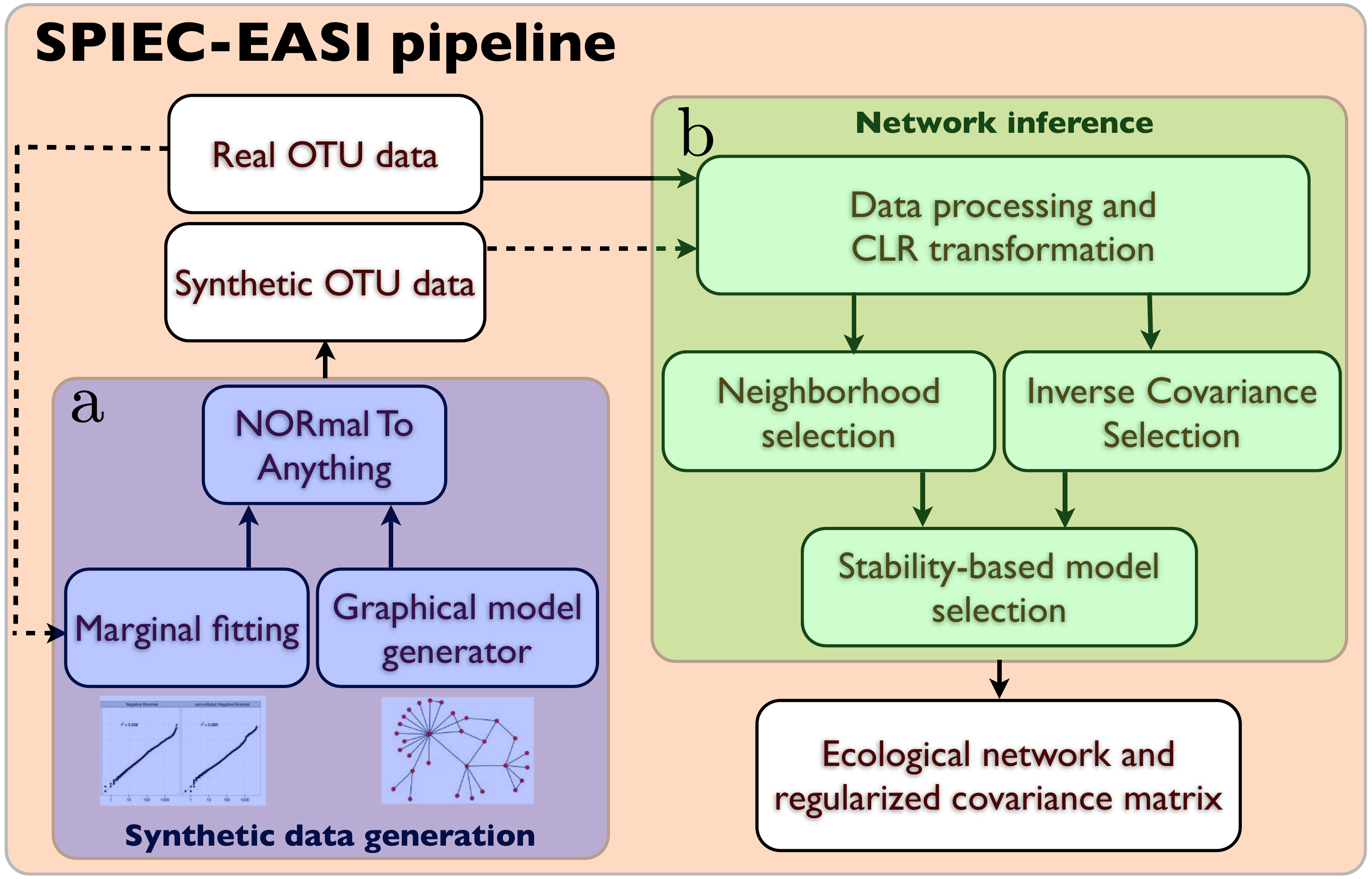}
\caption{\textbf{Workflow of the SPIEC-EASI pipeline}: The SPIEC-EASI pipeline consists of two independent parts for \textbf{a}) synthetic data generation and \textbf{b}) network inference. \textbf{a}) Synthetic data generation requires an OTU count table and a user-selected network topology. Internally, the parameters of a statistical distribution (the zero-inflated Negative binomial model is suggested) are fit to the OTU marginals of the real data, and are combined with the randomly-generated network in the Normal to Anything (NORTA) approach to generate correlated count data. \textbf{b}) Network inference proceeds in three stages on synthetic or real OTU count data: First, data is pre-procssed and centered log-ratio (CLR) transformed to ensure compositional robustness. Next, the user selects one of two graphical model inference procedures: 1) Neighborhood selection (the MB method) or 2) inverse covariance selection (the glasso method). SPIEC-EASI network inference assumes that the underlying network is sparse. We infer the correct model sparseness by the Stability Approach to Regularization Selection (StARS), which involves random subsampling of the dataset to find a network with low variability in the selected set of edges. SPIEC-EASI outputs include an ecological network (from the non-zero entries of the inverse covariance network) and an invertible covariance matrix. If the network was inferred from synthetic data, it can be compared with the input network to assess inference quality.}
\label{fig:spieceasi}
\end{figure}

SPIEC-EASI comprises both an inference and a synthetic data generation module. Figure \ref{fig:spieceasi} summarizes the key components of the pipeline. In this section, we introduce all statistical and computational aspects of the inference scheme and then describe our approach for generating realistic synthetic datasets.
\subsection*{Data processing and transformation of standard OTU count data}

For this discussion, a table of OTU count data, typical output of 16S rRNA gene sequencing data curation pipelines (e.g., mothur \cite{Schloss:2009}, QIIME \cite{LaRosa2012}) are given. The OTU data are stored in a matrix $\mathrm{W}\in\mathbb{N}_{0}^{n\times p}$ where $w^{(j)}=[w^{(j)}_{1},w^{(j)}_{2},\ldots,w^{(j)}_{p}]$  denotes the $p$-dimensional row vector of OTU counts from the $j^{th}$ sample, $j=1,\dots,n$, with total cumulative count $m^{(j)}=\underset{i=1}{\overset{p}{\sum}}w^{(j)}_{i}$; $\mathbb{N}_{0}$ denotes the set of natural numbers $\{0,1,2, \ldots\}$.
As described above, to account for sampling biases, microbiome data is typically transformed by normalizing the raw count data $w^{(j)}$ with respect to the total count $m^{(j)}$ of the sample \cite{Lee:2014}. We thus arrive at vectors of relative abundances or compositions $x^{(j)}=[\frac{w^{(j)}_{1}}{m^{(j)}},\frac{w^{(j)}_{2}}{m^{(j)}},\ldots,\frac{w^{(j)}_{p}}{m^{(j)}}]$ for sample $j$. Due to this normalization OTU abundances are no longer independent, and the sample space of this p-part composition $x^{(j)}$ is not the unconstrained Euclidean space but the $p$-dimensional unit simplex $\mathbb{S}^p \doteq \{ x \, | x_i>0, \sum_{i=1}^p x_i = 1\}$. Thus, OTU compositions from n samples are constrained to lie in the unit simplex, $\mathrm{X}\in\mathbb{S}^{n\times p}$.
This restriction of the data to the simplex prohibits the application of standard statistical analysis techniques, such as linear regression or empirical covariance estimation. Covariance matrices of compositional data exhibit, for instance, a negative bias due to closure effects.

Major advances in the statistical analysis of compositional data were achieved by Aitchison in the 1980's \cite{Aitchison1981,Aitchison1986}. Rather than considering compositions in the simplex, Aitchison proposed log-ratios, $\log[\frac{x_{i}}{x_{j}}]$, as a basis for studying compositional data. The simple equivalence $\log[\frac{x_{i}}{x_{j}}]=\log[\frac{w_{i}/m}{w_{j}/m}]=\log[\frac{w_{i}}{w_{j}}]$ implies that statistical inferences drawn from analysis of log-ratios of compositions are equivalent to those that could be drawn from the log-ratios of the unobserved absolute measurements, also termed the \textit{basis}. 

Aitchison also proposed several statistically equivalent log-ratio transformations to remove the unit-sum constraint of compositional data \cite{Aitchison1981}. Here we apply the centered log-ratio (clr) transform:
\begin{equation}
z \doteq \mathrm{clr}(x) =  [\log(x_{1}/g(x)),...,\log(x_{p}/g(x)] \\
=  [\log(w_{1}/g(w)),...,\log(w_{p}/g(w))] \label{eq:clr_w} \, 
\end{equation}
\label{eq:clr_x}
where $g(x)=\left[\underset{i=1}{\overset{p}{\prod}}x_{i}\right]^{1/p}$ is the geometric mean of the composition vector. The clr transform is symmetric and isometric with respect to the component parts. The resulting vector $z$ is constrained to a zero sum. The clr transform maps the data from the unit simplex to a $p-1$-dimensional Euclidean space, and the corresponding population covariance matrix $\Gamma =  \mathrm{Cov\left[clr(X)\right]} \in \mathbb{R}^{p \times p}$ is also singular \cite{Aitchison1981}. The covariance matrix $\Gamma$ is related to the population covariance of the log-transformed absolute abundances $\Omega =  \mathrm{Cov\left[\mathrm{log}W\right]}$
via the relationship \cite{Aitchison1986}:
\begin{equation}
\Gamma  =  \mbox{\ensuremath{\mathrm{G\Omega G}}}
\label{eq:clrcov}
\end{equation}
where $\mathrm{G} = \mathrm{I}_{p}-\frac{1}{p}\mathrm{J}\label{eq:G}$, $\mathrm{I}_{p}$ is the $p$-dimensional identity matrix, and $\mbox{\ensuremath{\mathrm{J}}} = [j_{1}, j_{2}, \ldots, j_{i}, \ldots ,j_{p}]$, $j_i = [1,1, \ldots, 1]$ the $p$-dimensional all-ones vector. For high-dimensional data, $p>>0$, the matrix $\mathrm{G}$ is close to the identity matrix, and thus we can assume that a finite sample estimator $\hat \Gamma$ of $\Gamma$ serves as a good approximation of $\hat \Omega$. This approximation serves as the basis of our network inference scheme. 
Finally, because real-world OTU data often contain samples with a zero count for low-abundance OTUs, we add a unit pseudo count to the original count data to avoid numerical problems with the clr transform.



\subsection*{Inference of microbial associations from microbial abundance datasets}
Our key objective is to learn a network of pairwise taxon-taxon associations (putative interactions) from clr-transformed microbiome compositions $Z \in \mathbb{R}^{n \times p}$. We represent the network as an undirected, weighted graph $\mathcal{G} = (V,E)$, where the vertex set $V = \{v_1,\ldots,v_p\}$ represents the $p$ taxa (e.g., OTUs) and the edge set $E \subset V \times V$ the possible associations among them. Our formal approach is to learn a probabilistic graphical model \cite{Koller:2009} (i) that is consistent with the observed data and (ii) for which the (unknown) graph $\mathcal{G}$ encodes the conditional dependence structure between the random variables (in our case, the observed taxa). Graphical models over undirected graphs (also known as Markov networks or Markov Random Fields) have a straightforward distributional interpretation when the data are drawn from a probability distribution $\pi(x)$ that belongs to an exponential family \cite{Lauritzen:1996,Wainwright:2008}. For example, when the data are drawn from a multivariate normal distribution $\pi(x) = \mathcal{N}(x | \mu,\Sigma)$ with mean $\mu$ and covariance $\Sigma$, the non-zero elements of the off-diagonal entries of the inverse covariance matrix $\Theta = \Sigma^{-1}$, also termed the precision matrix, defines the adjacency matrix of the graph $\mathcal{G}$ and thus describes the factorization of the normal distribution into conditionally dependent components \cite{Koller:2009}. Conversely, if and only if an entry in $\Theta$: $\Theta_{i,j}=0$, then the two variables are conditionally independent, and there is no edge between $v_i$ and $v_j$ in $\mathcal{G}$. We seek to estimate the inverse covariance matrix from the data, thereby inferring associations based on conditional independence. This is fundamentally distinct from SparCC and CCREPE (see Table \ref{methtable}), which essentially estimate pairwise correlations (though other pairwise metrics could be considered for CCREPE). We highlight this key difference in Figure \ref{toydata_all}. For an intuitive introduction to graphical models in the context of biological networks see B\"uhlmann  \textit{et. al}, 2014 \cite{Buhlmann:2014}.

Inferring the exact underlying graph structure in the presence of a finite amount of samples is, in general, intractable. However, two types of statistical inference procedures have been useful in high-dimensional statistics due to their provable performance guarantees under assumptions about the sample size $n$, dimensionality $p$, underlying graph properties, and the generating distribution \cite{Ravikumar:2010,Ravikumar2011}. The first approach, neighborhood selection \cite{Meinshausen2006,Ravikumar:2010}, aims at reconstructing the graph on a node-by-node basis where, for each node, a penalized regression problem is solved. The second approach is the penalized maximum likelihood method \cite{Banerjee2008,Friedman2008}, where the entire graph is reconstructed by solving a global optimization problem, the so-called covariance selection problem \cite{Dempster:1972}. The key advantages of these approaches are that (i) their underlying inference procedures can be formulated as convex (and hence tractable) optimization problems, and (ii) they are applicable even in the underdetermined regime $(p>n)$, provided that certain structural assumptions about the underlying graph hold. One assumption is that the true underlying graph is reasonably sparse, e.g., that the number of taxon-taxon associations scales linearly with the number of measured taxa. 

\paragraph{Graphical model inference.} The SPIEC-EASI pipeline comprises two types of inference schemes, neighborhood and covariance selection. The neighborhood selection framework, introduced by \textbf{M}einshausen and \textbf{B}\"uhlmann \cite{Meinshausen2006} and thus often referred as the MB method, tackles graph inference by solving $p$ regularized linear regression problems, leading to local conditional independence structure predictions for each node. Let us denote the $i^{th}$ column of the data matrix $Z$ by $Z^{i}\in\mathbb{R}^{n}$ and the remaining columns by $Z^{\neg i}\in\mathbb{R}^{n\times p-1}$. For each node $v_i$, we solve the following convex problem:
\begin{equation}
\hat \beta^{i,\lambda} =\underset{{\scriptstyle \beta \in \mathbb{R}^{p-1}}}{\arg\min} \left(\frac{1}{n} \|Z^{i}-Z^{\neg i}\beta \|^2 + \lambda \| \beta \|_1 \right) \, ,
\label{eq:mb}
\end{equation}
where $\| a \|_1 = \sum_{i=1}^{p-1} |a_i|$ denotes the L1 norm, and $\lambda\ge0$ is a scalar tuning parameter. This so-called LASSO problem \cite{Tibshirani1996} aims at balancing the least-square fit and the number of necessary predictors (the non-zero components $\beta_j$ of $\beta$) by tuning the $\lambda$ parameter. We define the local neighborhood of a node $v_i$ as $N^\lambda_i = \{ j \subset \{1,\ldots p \} \setminus i : \hat \beta^{i,\lambda}_j \neq 0 \}$. The final edge set $E$ of $\mathcal{G}$ can be defined via the intersection or the union operation of the local neighborhoods. An edge $e_{i,j}$ between node $v_i$ and $v_j$ exists if $j \in N^\lambda_i \cap i \in N^\lambda_j$ or $j \in N^\lambda_i \cup i \in N^\lambda_j$. For edges in the set $j \in N^\lambda_i \cap i \in N^\lambda_j$, the edge weights, $e_{i,j}$ and $e_{j,i}$, are estimated using the average of the two corresponding $\beta$ entries. From a theoretical point of view, both edge selection choices are asymptotically consistent under certain technical assumptions \cite{Meinshausen2006}. The choice of the $\lambda$ parameter controls the sparsity of the local neighborhood, which requires tuning \cite{Lederer:2014}. We present our parameter selection strategy at the end of this section.

The second inference approach, (inverse) covariance selection, relies on the following penalized maximum likelihood approach. In the standard Gaussian setting, the related convex optimization problem reads:
\begin{equation}
\hat{\Theta} =  \underset{{\scriptstyle \Theta \in \text{PD}}}{\arg\min} \left( -\log\det(\Theta)+\mathrm{tr}(\Theta \hat \Sigma)+\lambda\left\Vert \Theta \right\Vert _{1} \right) \, ,
\label{eq:spic}
\end{equation}
where $\text{PD}$ denotes the set of symmetric positive definite matrices $\{A: x^T A x >0 $, $\forall x \in \mathbb{R}^p\}$, $\hat \Sigma$ the empirical covariance estimate, $\| \cdot \|_1$ the element-wise L1 norm, and $\lambda \ge 0$ a scalar tuning parameter. For $\lambda=0$, the expression is identical to the maximum likelihood estimate of a normal distribution $\mathcal{N}(x | 0,\Sigma)$. For non-zero $\lambda$, the objective function (also referred as the graphical Lasso \cite{Friedman2008}) encourages sparsity of the underlying precision matrix $\Theta$. The non-zero, off-diagonal entries in $\Theta$ define the adjacency matrix of the interaction graph $\mathcal{G}$ which, similar to MB, depends on the proper choice of the penalty parameter $\lambda$. Originally, this estimator was shown to have theoretical guarantees on consistency and recovery only under normality assumptions \cite{Lam:2009}. However, recent theoretical \cite{Ravikumar2011,Loh:2013} work shows that distributional assumptions can be considerably relaxed, and the estimator is applicable to a larger class of problems, including inference on discrete (count) data. In addition, nonparametric approaches, such as sparse additive models, can be used to ``gaussianize" the data prior to network inference \cite{Liu:2009}. We thus propose the following estimator for inferring microbial ecological associations. Given clr-transformed OTU data $Z \in \mathbb{R}^{n \times p}$, we propose the modified optimization problem:  
\begin{equation}
{\hat\Omega^{-1}} =  \underset{{\scriptstyle \Omega^{-1} \in PD}}{\arg\min} \left( -\log\det(\Omega^{-1})+\mathrm{tr}(\Omega^{-1} \hat \Gamma)+\lambda\left\Vert \Omega^{-1} \right\Vert _{1} \right) \, ,
\label{eq:spicOTU}
\end{equation}
where $\hat \Gamma$ is the empirical covariance estimate of $Z$, and $\Omega^{-1}$ is the inverse covariance (or precision matrix) of the underlying (unknown) basis. As stated above,  $\hat \Gamma$ will be a good approximation for the basis covariance matrix $\hat \Omega$ because $p>>0$. The resulting solution is constrained to the set of PD matrices, ensuring that the penalized estimator has full rank $p$. The non-zero off-diagonal entries of the estimated matrix $\Omega^{-1}$ define the inferred network $\mathcal{G}$, and their values are the signed edge weights of the graph. To reduce the variance of the estimate, the covariance matrix  $\hat \Gamma$ can also be replaced by the empirical correlation matrix $\hat R = D \hat \Gamma D$, where $D$ is a diagonal matrix that contains the inverse of the estimated element-wise standard deviations.

The covariance selection approach has two advantages over the neighborhood selection framework. First, we obtain unique weights associated with each edge in the network. No averaging or subsequent edge selection is necessary. Second, the covariance selection framework provides invertible precision and covariance matrix estimates that can be used in further downstream microbiome analysis tasks, such as regression and discriminant analysis \cite{Lee:2014}.

\paragraph{Model selection.} For both neighborhood and covariance selection, the tuning parameter $\lambda \in [0, \lambda_{\max}]$ controls the sparsity of the final model. Rather than inferring a single graphical model, both methods produce a $\lambda$-dependent solution path with the complete and the empty graph as extreme networks. A number of model selection criteria, such as Bayesian Information Criteria \cite{Yuan:2007} and resampling schemes \cite{hastie2009elements}, have been used. Here we use a popular model selection scheme known as the \textbf{St}ability \textbf{A}pproach to \textbf{R}egularization \textbf{S}election (StARS) \cite{Liu2010}. This method repeatedly takes random subsamples ($80\%$ in the standard setting) of the data and estimates the entire graph solution path based on this subsample. For each subsample, the $\lambda$-dependent incidence frequencies of individual edges are retained, and a measure of overall edge stability is calculated. StARS selects the $\lambda$ value at which subsampled non-empty graphs are the least variable (most stable) in terms of edge incidences. For the selected graph, the observed edge frequencies indicate the reproducibility, and likely the predictive power, and are used to rank edges according to confidence.

\paragraph{Theoretical and computational aspects.} Learning microbial graphical models with neighborhood or inverse covariance selection schemes has important theoretical and practical advantages over current methods. A wealth of theoretical results are available that characterize conditions for asymptotic and finite sample guarantees for the estimated networks \cite{Meinshausen2006,Lam:2009,Yuan:2007,Ravikumar:2010,Ravikumar2011}. Under certain model assumptions, the number of samples $n$ necessary to infer the true topology of the graph in the neighborhood selection framework is known to scale as $n = O(d^3 \log(p))$, where $d$ is the maximum vertex (or node) degree of the underlying graph (i.e. the maximum size of any local neighborhood). Additional assumptions on the sample covariance matrices reduce the scaling to $n = O(d^2 \log(p))$ \cite{Ravikumar:2010}. This implies that graph recovery and precision matrix estimation is indeed possible even in the $p>>n$ regime, and that the underlying graph topology strongly impacts edge recovery. The latter observation means that, even if the number of interactions $e$ is constant, graphs with large hub nodes, perhaps representing keystone species in microbial networks, or, more generally, scale-free graphs with, a few highly connected nodes, will be more difficult to recover than networks with evenly distributed neighborhoods. In addition to these theoretical results, a second advantage is that well-established, efficient, and scalable implementations are available to infer microbial ecological networks from OTU data in practice. Thus, SPIEC-EASI methods will efficiently scale as microbiome dataset dimensions grow (e.g., due to technological advances that increase the number of OTUs detected per sample). The SPIEC-EASI inference engine relies on the R package \textit{huge} \cite{Zhao2012}, which includes algorithms to solve neighborhood and covariance selection problems \cite{Meinshausen2006,Friedman2008}, as well as the StARS model selection.

\subsection*{Generation of synthetic microbial abundance datasets}
Estimating the absolute and comparative performance of network inference schemes from biological data remains a fundamental challenge in biology. In the context of gene regulatory network inference, recent community-wide efforts, such as the DREAM (Dialogue for Reverse Engineering Assessments and Methods) Challenges (\url{http://www.the-dream-project.org/}), have considerably advanced our understanding about feasibility, accuracy, and applicability of a large number of developed methods. In the DREAM challenges, both real data from ``gold standard" regulatory networks (e.g., networks where the true topology is known from independent experimental evidence) and realistic in-silico data (using, e.g., the GeneNetWeaver pipeline \cite{Schaffter2011}) are included. In the context of microbiome data and microbial ecological networks, neither a gold standard nor a realistic synthetic data generator exist. SPIEC-EASI is accompanied by a set of computational tools that allow the generation of realistic synthetic OTU data. As outlined in Figure \ref{fig:spieceasi}, real taxa count data serve as input to SPIEC-EASI's synthetic data generation pipeline. The pipeline enables one to: (i) fit the marginal distributions of the count data to a parametric statistical model and (ii) specify the underlying graphical model architecture (e.g., scale-free). 

\begin{figure}
\centering
\includegraphics[scale=0.42]{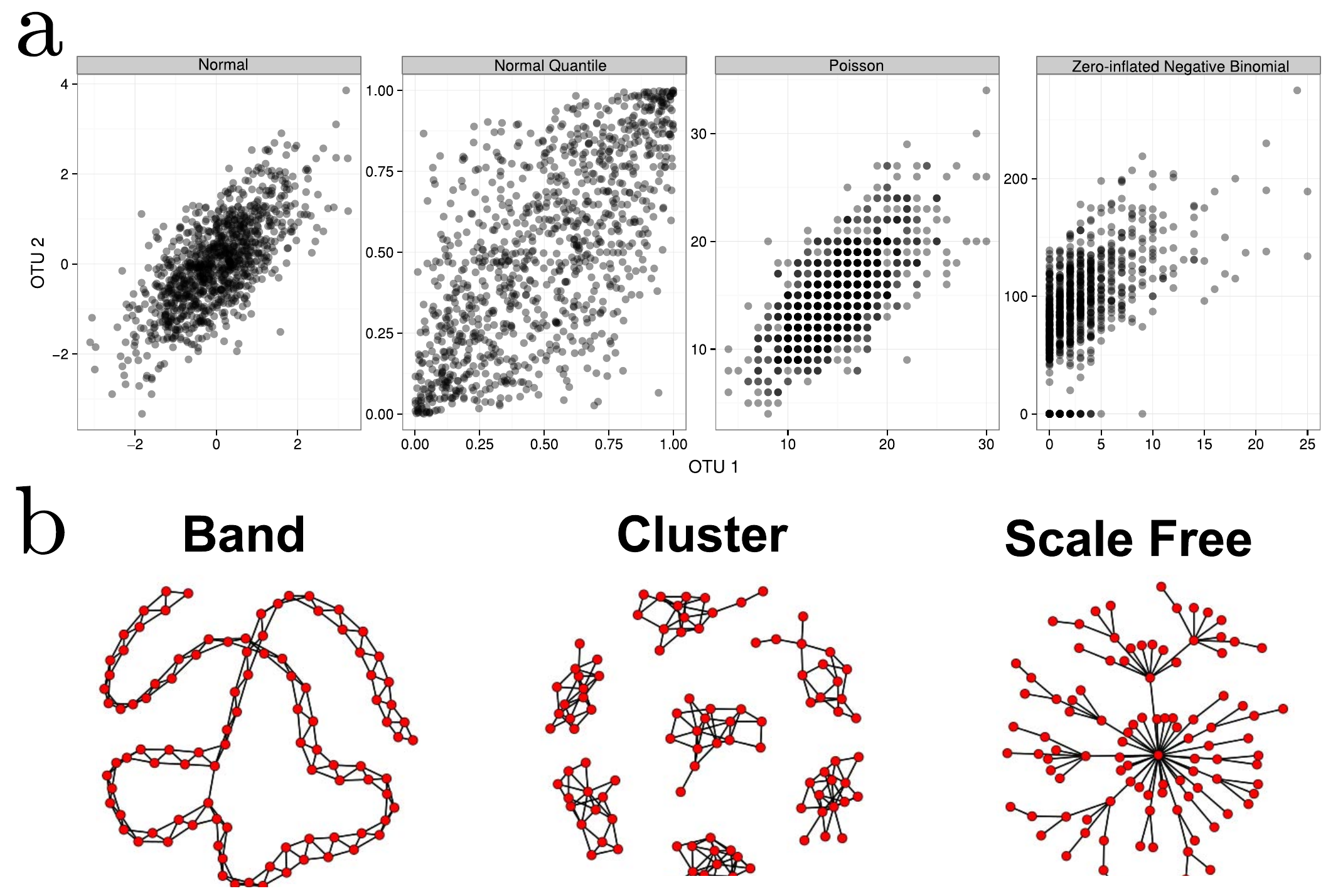}
\caption{\textbf{a}) \textbf{Bivariate illustration of the NorTA approach}. First normal data, incorporating the target correlation structure, is generated. Uniform data are then generated for each margin via the normal density function. These is then converted to an arbitrary marginal distribution (Poisson and Zero-inflated Negative Binomial shown as examples) via its quantile function. To generate realistic synthetic data, parameters for these margins are fit to real data. \textbf{b}) Examples of band-like, cluster, and scale-free network topologies}
\label{Fig_NORTA}
\end{figure}

\paragraph{The NorTA approach.}
The parametric statistical model and network topology are then combined in the 'Normal To Anything' (NorTA) \cite{Nelsen1999} approach to generate synthetic OTU data that resemble real measurements from microbial communities but with user-specified network topologies. NorTA \cite{Nelsen1999} is an approximate technique to generate arbitrary continuous and discrete multivariate distributions, given (1) a target correlation structure $R$ with entries $\rho_{i,j}$ and (2) a target univariate marginal distribution $U_{i}$. To achieve this task, NorTA relies on normal-copula functions \cite{Nelsen1999,Madsen2007,Cario1997}. A $n\times p$ matrix of data $\mathrm{U}$ is sampled from a normal distribution with zero mean and a $p\times p$ correlation matrix $\mathrm{R_\text{N}}$. For each marginal $U_{i}$, the Normal cumulative distribution function (CDF) is transformed to the target distribution via its inverse CDF. For any target distribution $P$ with CDF $\Xi$, we can thus generate multivariate correlated data via
\begin{equation}
\mathrm{\mathrm{U}}_{P_{i}}=\Xi^{-1}(\Phi(\mathrm{\mathrm{U}}_{N_{i}})) \, ,
\end{equation}
where $\mathrm{\mathrm{U}}_{N}\sim \mathcal{N}(0,\mathrm{R_\text{N}})$ and $\Phi(\mathrm{U})={\displaystyle \intop}_{-\infty}^{U}\frac{1}{\sqrt{2\sigma^{2}}}e^{\frac{-u^{2}}{2}}du$,
the CDF of a univariate normal. In Figure \ref{Fig_NORTA}a, we illustrate this process for bivariate Poisson and negative binomial data ($n=1000$ and correlation $\rho_{ij}=0.7$). 
In the original NorTA approach, an element-wise monotone transformation $c_\text{U}(\cdot)$ with $R_\text{N} = c_\text{U}(R)$ is applied to account for slight differences in correlation structure between normal and target distribution samples \cite{Nelsen1999}. Here we neglect this transformation step because we observe that the log-transformed data from exponential count distributions, such as the Poisson and Negative binomial, are already close to $\mathrm{R}$, provided that the mean is greater than one, particularly when the counts data are log-transformed (Appendix \ref{sec:norta_recov}). 
In practice, SPIEC-EASI relies on routines from base R and VGAM packages \cite{RDevelopmentCoreTeam2011, Yee2007} to estimate the uniform quantiles of the normal data and to fit the desired CDF with estimated parameters. 

\paragraph{Fitting marginal distribution to real OTU data.}
Prior to fitting marginal distributions to real data, several commonly used pre-processing steps are applied. For any given OTU abundance table of size $n \times K$, we first select $p$ non-zero columns. To account for experimental differences in sample sequencing, we then normalize samples to a median sequencing depth by multiplying all counts by the ratio of minimum desirable sampling depth to the total sum of counts in that sample and rounding to the nearest whole number, which is preferable to rarefaction \cite{McMurdie2014}. 
These filtered and sequencing-depth-normalized data serve as the marginal counts, which are fit to a parametric distribution $U_i$ and used as input to the NorTA approach. The concrete target marginal distribution depends on the actual microbiome dataset. For gut microbiome data (e.g. from HMP or APG), the zero-inflated Negative Binomial (ziNB) distribution is a good choice, as it accounts for both overdispersion  \cite{McMurdie2014,la2012hypothesis} and the preponderance of zero-count data points in microbial count datasets. The fitting procedure is done within a maximum likelihood framework. The corresponding optimization problem is solved with the Quasi-Newton methods with box-constraints, as implemented in the optim function in R \cite{RDevelopmentCoreTeam2011}. In Appendix \ref{sec:amgut_fit}, we use quantile-quantile plots to compare ziNB to several other candidate distributions (e.g., lognormal, Poisson, NB) and show that ziNB has superior fit.

\paragraph{Generation of network topologies and correlation matrices.}
Under normality assumption, the non-zero pattern of the precision matrix corresponds to the adjacency matrix of the underlying undirected graph. We use this property to generate target covariance (correlation) matrices originating from different graph topologies. The pipeline to generate a network structure for simulated data proceeds in three steps: (i) Generate an undirected graph, in the form of an adjacency matrix, with a desired topology and sparsity, (ii) convert the adjacency matrix to a positive-definite precision matrix  by assigning positive and negative edge weights and appropriate diagonal entries, and (iii) invert $\Theta$ and convert the resulting covariance matrix $\Sigma$ to a correlation matrix ($R = D \Sigma D$, where, $D$ is a diagonal matrix with diagonal entries $1/\sqrt{\sigma_i}$). 

Among many potential graph structures, we focus on three representative network structures:
band-like, cluster, and scale-free graphs (see Figure \ref{Fig_NORTA}b for graphical examples). Maximum network degree strongly impacts network recovery, and thus our choice of network topologies spans a range of maximum degrees (band $<$ cluster $<$ scale-free). In addition, cluster and scale-free lend themselves to hypothetical ecological scenarios. Cluster graphs may be seen as archetypal models for microbial communities that populate different disjoint niches (clusters) and have only few associations across niches. Scale-free graphs, ubiquitous in many other facets of network biology (such as gene regulatory, protein-protein and social networks), serve as a baseline model for a microbial community that comprises (1) a few ``keystone" species (hub nodes with many partners) that are essential for coordinating/stabilizing the community and (2) many dependent species that are sparsely connected to each other. 
The sparsity of the networks is controlled by the number of edges, $e < p(p-1)/2$, in the graph. The topologies are generated according to the following algorithms, starting with an empty $p \times p$ adjacency matrix:

\begin{enumerate}
\item \textbf{Band:} A band-type network consists of a chain of nodes that connect only their nearest neighbors. Let $e=e_{used}+e_{available}$, the number of edges already used and available, respectively. Fill the next available off-diagonal vector with edges if and only if $e_{available}\geq$
number of elements in the off-diagonal.
\item \textbf{Cluster:}
A cluster network comprises $h$ independent groups of randomly connected nodes. For given $p$ and $e$ we divide the set of nodes into $h$ components of (approximately) identical size and set the number of edges in each component to $e_\text{comp}=e/h$. For each component, we generate a random (Erd\"os-Renyi) graph for which we randomly assign an edge between two nodes in the cluster with probability $p=\frac{e_\text{comp}}{h(h-1)/2}.$
\item \textbf{Scale-free:} The distribution of degrees, the number of edges per node, in a scale-free network is described by a power law, implying that the central node or nodes (potentially keystone species in an ecological network) have proportionally more connections. We use the standard preferential attachment scheme \cite{Barabasi1999} until $p-1$ edges are exhausted. 
\end{enumerate}
After generation of these standard adjacency matrices, we randomly remove or add edges until the adjacency matrix has exactly $e$ edges. All schemes generate symmetric adjacency matrices that describes a graph, with entries of 1 if an edge exists and 0 otherwise. 

From the adjacency matrices, we generate precision matrices by uniformly sampling non-zero entries $\Theta_{i,j} \in [-\Theta_\text{max},-\Theta_\text{min}] \cup [\Theta_\text{min},\Theta_\text{max}]$, where  
$\Theta_\text{min},\Theta_\text{max} > 0$ are model parameters and describe the strength of the conditional dependence among the nodes. To ensure that the precision matrix is positive definite with tunable condition number $\kappa = \text{cond}(\Theta)$, we scale the diagonal entries $\Theta_{i,i}$ by a constant $c$ using binary search. The precision matrix $\Theta$ is then converted to a correlation matrix $R$ to be used as input to the NorTA approach.

\section*{Results}
\label{sec:results}
\subsection*{Network inference on synthetic microbiome data}

Given that no large-scale experimentally validated microbial ecological network exists, we use SPIEC-EASI's data generator capabilities to synthesize data whose OTU count distributions faithfully resemble microbiome count data. By varying parameters known to influence network recovery (network topology, association strength, sample number) and quantifying performance on resulting networks, we rigorously assess SPIEC-EASI inference relative to state-of-the-art inference methods, SparCC \cite{Friedman2012} and CCREPE \cite{Gevers:2014}, as well as standard Pearson correlation.

\subsubsection*{Benchmark setup}
We modeled the synthetic datasets on American Gut Project data using SPIEC-EASI's data generation module. The count data, accessed February, 2014 at \url{www.microbio.me/qiime} and available at \url{https://github.com/zdk123/SpiecEasi/tree/master/inst/extdata}, come from two sampling rounds and comprise several thousand OTUs. Round 1 data contains $n_1 = 304$, and Round 2 data contains $n_2=254$ samples. As filtering steps, OTUs were removed from the input data if present in fewer than $37\%$ of the samples, while samples were removed if total sequencing depth fell below the 1st quartile (10,800 sequence reads). Thus, we arrived at a total of $p=205$ distinct OTUs. We also generated smaller-dimensional datasets ($p=68$) with fewer zero counts by requiring that OTUs be present in $>60\%$ of the samples. We used Round 1 data and fit the $n_1$ count histograms to a ziNB distribution (for justification of this, see Appendix \ref{amgutfit}). The empirical effective number $n_\text{eff}$ is 13.5 for $p=205$ and 7.5 for $p=68$ data. The resulting parametrized marginal distributions served as input to NorTA. 


As described above, network topology is expected to influence network recovery; thus, we consider the three previously described topologies (band-like, cluster, and scale-free) as representative microbial networks. We hypothesize that any method that successfully infers the sets of associations underlying these archetypal networks from synthetic datasets is likely to also perform well in the context of true microbiomes, whose underlying network architecture is unknown but expected to be a mixture of these network types. For all networks, we fix the total number of edges $e$ to the respective number of OTUs $p$, and we analyze a medium ($p=68$) and a high-dimensional scenario ($p=205$). Microbial association strength is controlled by the range of values in off-diagonal entries in the precision matrices $\Theta$ and the condition number $\kappa = \text{cond}(\Theta)$. We use $\Theta_\text{min} = 2$ and $\Theta_\text{max} = 3$ with either condition number $\kappa =10$ or $100$. In this setting, $\kappa$ controls the spread of the absolute correlation values (and thus the strength of indirect associations) present in the synthetic data. The relationship between condition number and distribution of correlation is illustrated in Figure \ref{cond_corr}. For each network type and size, we generate 20 distinct instances. For each instance, we then use the NorTA approach to generate a maximum of $n=1360$ synthetic microbial count data samples. In Figure \ref{amgutfit_heatmap}, we highlight the fidelity between the target microbiome dataset and the synthetic datasets, especially in terms of dataset dimensions and OTU count distributions. To assess the effect of sample size on network recovery, we test methods on a range of sample sizes: $n=34,68,102,1360$. 

We compare SPIEC-EASI's covariance selection method (referred to as S-E(glasso)) and neighborhood selection method (referred to as S-E(MB)) to  SparCC and CCREPE, methods which were also designed to be robust to compositional artifacts. As a baseline reference, we also compared all methods to Pearson correlation, which is neither compositionally robust nor appropriate for estimating correlation in the under-determined regime. Both of these methods, however, infer interactions from correlations and do not consider the concept of conditional independence. We improved the runtime of the original SparCC implementation (available at \url{https://bitbucket.org/yonatanf/sparcc}) and include the updated code in our SPIEC-EASI package. The original SparCC package also includes a small benchmark test case available at \url{https://bitbucket.org/yonatanf/sparcc/src/9a1142c179f7/example}). The recovery performance results for S-E(MB), SparCC, and CCREPE for this test case can be found in Figure \ref{sparcc_benchmark}. The CCREPE implementation is downloaded from \url{http://www.bioconductor.org/packages/release/bioc/html/ccrepe.html}. 


\subsubsection*{Recovery of microbial networks}
\begin{figure}
	\centering
    \includegraphics[scale=.35]{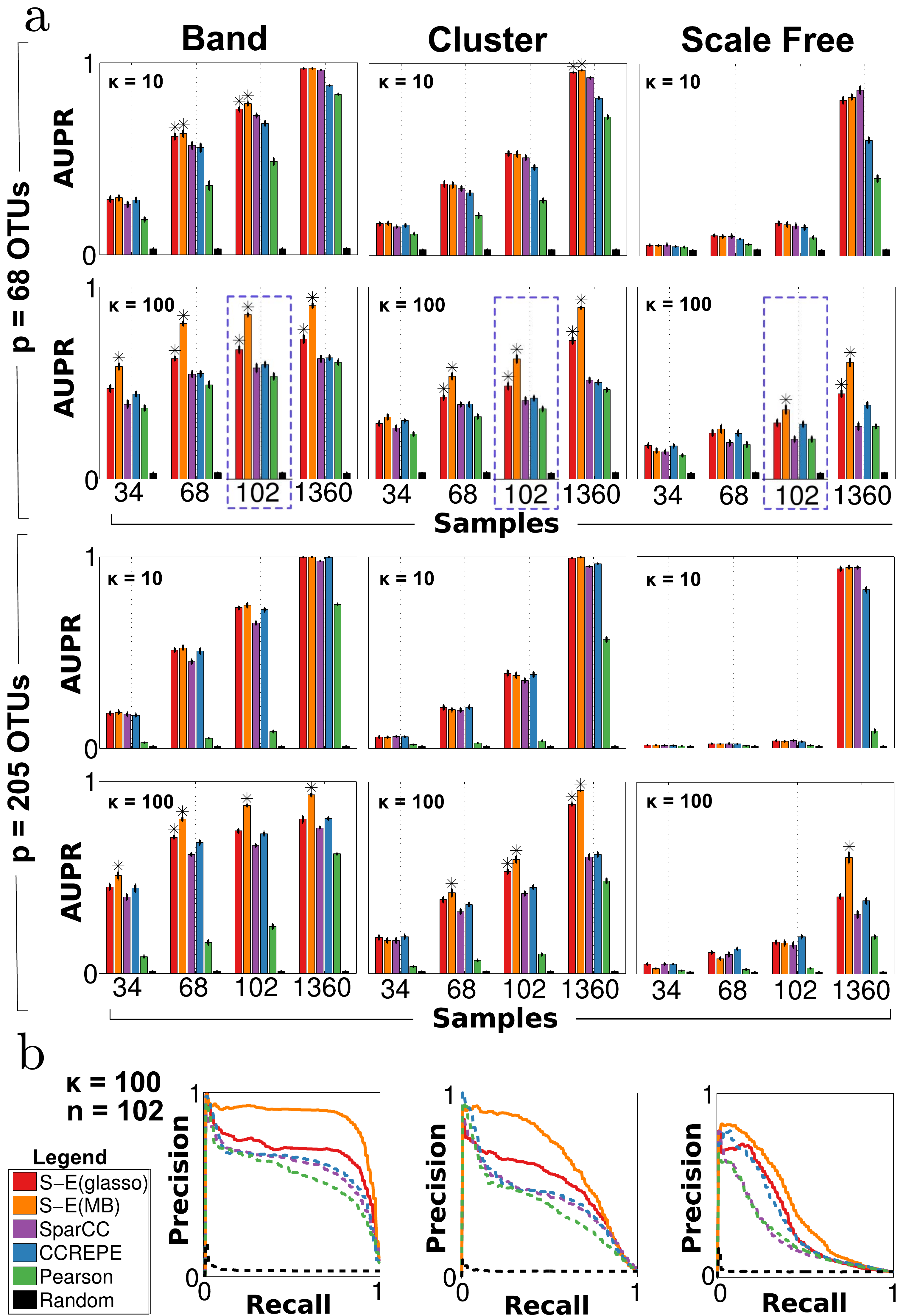}	
\caption{\textbf{Precision-recall performance on synthetic datasets}. \textbf{a}) Red=S-E(glasso), orange=S-E(MB), purple=SparCC, blue=CCREPE, green=Pearson correlation, black=random. Area under precision-recall (AUPR) vs. number of samples $n$ for different $\kappa$ values are depicted. Bars represent average over 20 synthetic datasets, and error bars represent standard error. Asterisks denote conditions under which SPIEC-EASI methods had significantly higher AUPR relative to all other control methods (P<0.05 for all one-sided T tests). \textbf{b}) Representative precision-recall curves for $p=68$, $n=102$, $\kappa=100$; solid and dashed lines denote SPIEC-EASI and control methods, respectively.}
\label{PRboth}
\end{figure}

To quantify each method's ability to recover the true underlying association network, we evaluated performance in terms of precision-recall (P-R) curves and area under P-R curves (AUPR). For each method, we ranked edge predictions according to confidence. For SparCC, CCREPE and Pearson correlation, edge predictions were ranked according to p-value. SPIEC-EASI edge predictions were ranked according to edge stability, inferred by StARS model selection step at the most stable tuning parameter $\lambda_\text{StARS}$. Figure \ref{PRboth} summarizes methods' performance on 960 independent synthetic datasets for a total of 48 conditions (4 samples sizes $\times$ 2 conditions numbers $\times$ 3 network topologies $\times$ 2 dimensions).

We observe the following key trends. First, the performance of all methods improves with increasing sample size. Under certain scenarios, even near-perfect recovery (AUPR $\approx1$) is possible in the large sample limit ($n=1360$). Second, all methods show a clear dependence on the network topology. Best performance is achieved for band graphs, followed by cluster and scale-free graphs. These results are consistent with theoretical results \cite{Ravikumar2011}, which show that the maximum node degree $d$ reduces the probability of correctly inferring network edges for fixed sample size (scale-free networks have highest maximum degree, followed by cluster and then band.) Third, for most scenarios, the SPIEC-EASI methods, particularly S-E(MB), perform as well or significantly better than all control methods. Standard Pearson correlation is outperformed by all methods that take the compositional nature of the data into account. Forth, in the large sample limit ($n=1360$), S-E(MB) is the only method that recovers a significant portion of edges under all tested scenarios (particularly scale-free networks). Additionally, we observe (Figure \ref{sparccfake}) that SPIEC-EASI performs well on synthetic data generated by the SparCC benchmark \cite{BitData, Friedman2012}.

The present results suggest that complete network recovery is likely an unrealistic goal for microbiome studies, given that most studies have at most hundreds of samples. In addition, the P-R curves are based on ranking predicted edges. To generate a final network, confidence-based criteria must be applied to select a final set of edges for network inclusion, and, to date, no optimal selection process exists. Nonetheless, if we focus on the set of high-confidence interactions (i.e. the top-ranked entries in the edge list), we see that S-E methods, particularly S-E(MB), can achieve very high precision for all network types (see Figure \ref{PRboth} for a representative P-R curve). 

Overall, these results suggest that S-E methods outperform current state-of-the-art methods in terms of network recovery under most tested scenarios, with S-E(MB) showing superior performance over S-E(glasso). 


\subsubsection*{Recovery of Global Network Properties}
Accurate recovery of global network properties (e.g., degree distribution, number of connected components, shortest path lengths) from taxa abundance data would help define the underlying topology of the ecological network (e.g., cluster versus scale-free) \cite{Barabasi1999,Deng:2012}. At this point, little is known about the underlying topology of microbial ecological networks, but, as elaborated in the \nameref{sec:discussion}, such information could be incorporated as a constraint into SPIEC-EASI's inference methods, thereby further improving prediction. Thus, we tested how well SPIEC-EASI and other methods recover global network properties from the synthetic datasets and evaluate whether these methods might be able to provide insight into global network architecture, perhaps even in the underdetermined regime, where the prediction of individual edges is less accurate. To control for the disparate means by which individual methods ranked edge confidences (e.g., stability for the SPIEC-EASI methods, estimation of p-values for  SparCC, Pearson and CCREPE), for each synthetic dataset and method, final networks were generated by selecting the top 205 predicted edges and comparing to the true synthetic network topologies for $p = 205$ OTUs.

\begin{figure}
\centering
\includegraphics[scale=.75]{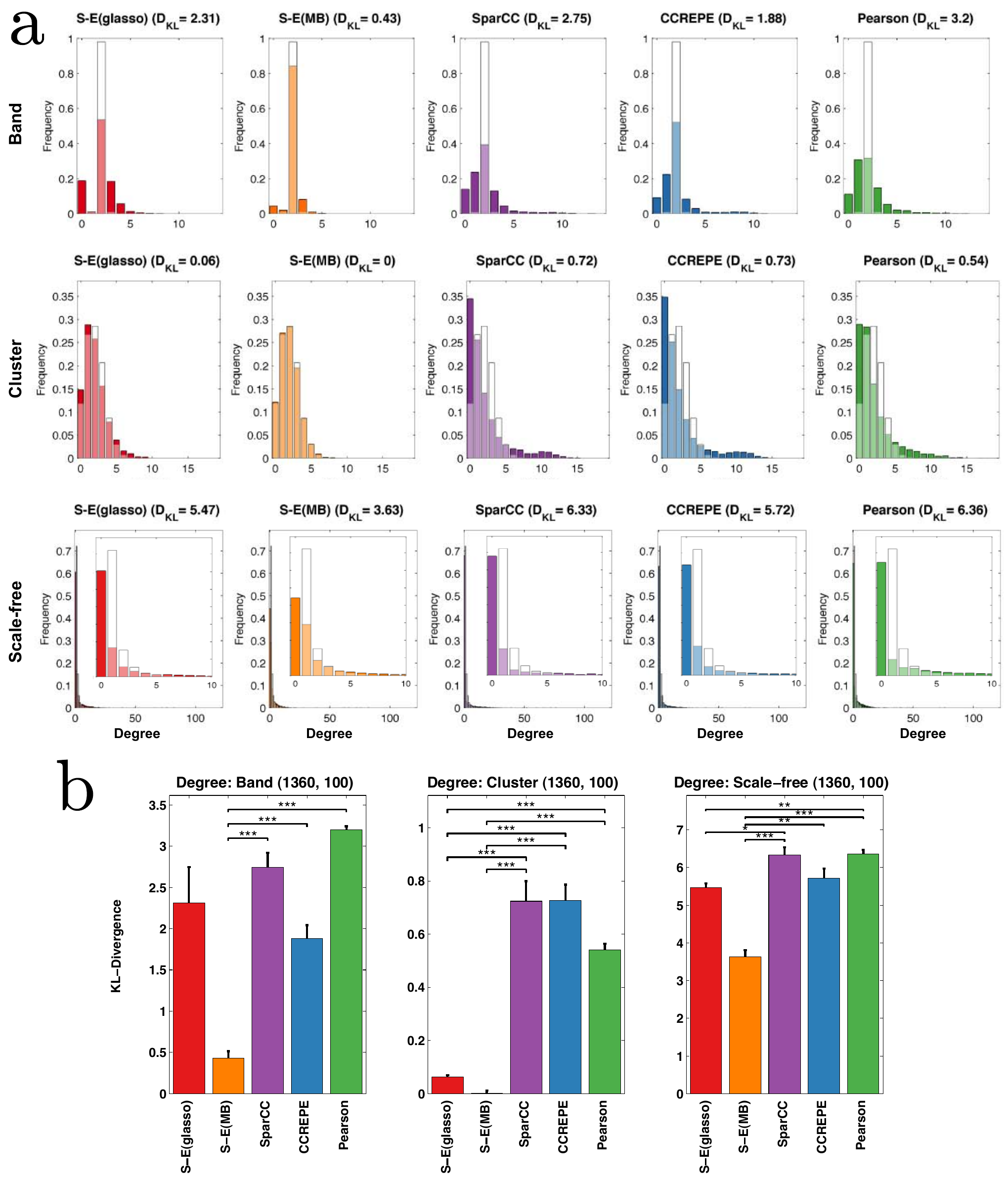}
\caption{\textbf{a}) Predicted degree distributions (colored) are overlaid with the true degree distribution (white) for $n = 1360$ samples, $p = 205$ OTUs, $\kappa = 100$. Lighter shades correspond to regions of overlap between predicted and true distributions. Dissimilarity between the distributions is measured by KL divergence, $D_\text{KL}$. \textbf{b}) Bars represent the average $D_\text{KL}$ over three independent sets of synthetic datasets (7 datasets per set); error bars represent standard error. Divergences were compared between S-E and control methods using one-sided T-tests; ***,**,* correspond to P<0.001, 0.01, and 0.05.}
\label{degree_hist}
\end{figure}
We first consider (node) degree distributions, where node degree is defined as the number of edges each node has. In Figure \ref{degree_hist}, we show the empirical degree distribution and the underlying ground truth for all methods and networks types, $n=1360$ and $\kappa=100$. Scale-free networks are characterized by exponential degree distributions, in which few nodes (e.g., hubs and, in our context, potential keystone taxa) have very high degree (e.g., interact with other taxa), while most nodes/taxa have few interactions. In contrast, nodes in cluster networks have relatively even degree, which depends on cluster size. In the ecological context, cluster networks would be consistent with niche communities that share few interactions with microbiota outside of one's niche community; this structure is also reflected in degree distributions. Using Kullback-Leibler (KL) divergence to measure the dissimilarity between methods' predicted degree distributions and the true degree distribution we see that S-E(MB) outperforms all other methods in recovering degree distributions (Figure \ref{degree_hist}). This performance improvement also holds for smaller samples sizes.

Another common topological feature is betweenness centrality, which, similar to degree, betweenness centrality can be used to gauge the relative importance of a node (e.g., taxon) to the (ecological) network. Betweenness centrality, as the fraction of shortest paths between all other nodes in the network that contain the given node, highlights central nodes. The distribution of nodes' betweenness centrality provides information about the network architecture (Figure \ref{hist_between}). Specifically, scale-free networks are expected to have a few nodes with very high betweenness centrality that connect most other nodes to each other; in scale-free networks, betweenness centrality can approach unity. For cluster networks, the maximum betweenness centrality is limited by the total number of independent clusters. In band networks, similar to scale-free, all nodes are connected; however, the degree is fixed and so the betweenness centrality distribution is roughly uniform from zero to one. For smaller sample sizes $n < 1360$, no method dominates. However, for the largest sample size, $n = 1360$, S-E(MB) is again significantly better than all other methods for five out of six conditions with the exception of scale-free networks $\kappa=10$, where SparCC recovery is best (Figure \ref{betweenness}).

We next consider distributions over graph geodesic distances. The geodesic distance is the length of the shortest path between two nodes. Given the existence of highly connected hubs in scale-free networks, geodesic distances for scale-free networks tend to be short, a feature that is described as the "small-world" property. Thus, the geodesic distributions in the scale-free network are a lot smaller than for the band and cluster networks (Figure \ref{geod_hist}). In recovery of geodesic distance distributions, S-E(MB) performs equivalently or significantly better than all other methods for scale-free networks as well as band graphs across all conditions. For cluster networks, the other methods generally outperform SPIEC-EASI methods for smaller sample sizes ($n < 1360$). In the large sample limit $n = 1360$, S-E(MB) has significantly better recovery of geodesic distance distributions relative to all control methods, even for cluster graphs (Figure \ref{geodesic}).

Finally, we analyzed the number and size of connected components in the inferred graphs. While all synthetic band and scale-free synthetic networks form a single connected component containing all nodes, cluster networks have a variable number of connected components. In terms of cluster number recovery, all methods predicted too many connected components. Overall, S-E methods had lower error rates for band and scale-free networks over all sample sizes. For high sample number ($n=1360$), S-E(MB) had significantly better recover of cluster size across all network types (Figure \ref{clustersize}), with nearly perfect recovery for cluster graphs ($D_{KL}=0$, Figure \ref{clustersize2}).

\subsection*{Inference of the American Gut network}
Thus far we have used the $n_1 = 304$ first-round AGP samples as a means to construct realistic synthetic microbiome data sets with SPIEC-EASI's data generation module.  In this section, we apply SPIEC-EASI inference methods to construct ecological association networks from the AGP data directly. To do this, we first filter out rare OTUs by selecting only the top 205 OTUs (to match the dimensionality of the synthetic data) in the combined AGP dataset (by frequency of presence) before adding a pseudo-count and total-sum normalization. Although there is no independent means to assess the accuracy of these hypothetical networks, we can assess their reproducibility and consistency. For each method, we first infer a single representative network of taxon-taxon interactions from Round 1 AGP abundance data. For SPIEC-EASI, the StARS model selection approach is used to select the final model network. For SparCC, we use a threshold $\rho_t=0.35$ to construct a relevance network from the SparCC-inferred correlation matrix; i.e. an edge between nodes $v_i$, $v_j$ is present in the SparCC network if $|\rho_{i,j}| > \rho_\text{t}$ \cite{Friedman2012}. Similarly, we use a q-value cut-off of $10^{-24}$ to create an interaction network from CCREPE-corrected significance scores of Pearson's correlation coefficient \cite{Gevers:2014}. For each method, we thus arrive at a reference network that can be considered the hypothetical gold standard. We then use the $n_2=254$ Round 2 AGP samples as an independent test set and learn a new  model network from these data alone. We measure consistency between the two network models by computing the Hamming distance between the reference and new network models, i.e., the difference between the upper triangular part of the two adjacency matrices. For the present data, the Hamming distance can vary between $p(p-1) / 2 = 20910$ (no edges in common) and a minimum of $0$ for identical networks. Confidence intervals for Hamming distances can be obtained by combining Round 1 and 2 samples into a unified dataset, repeatedly subsampling these data into two disjoint groups of size $n_1$ and $n_2$, and repeating the entire inference procedure.

\begin{figure}
\centering
\includegraphics[scale=0.52]{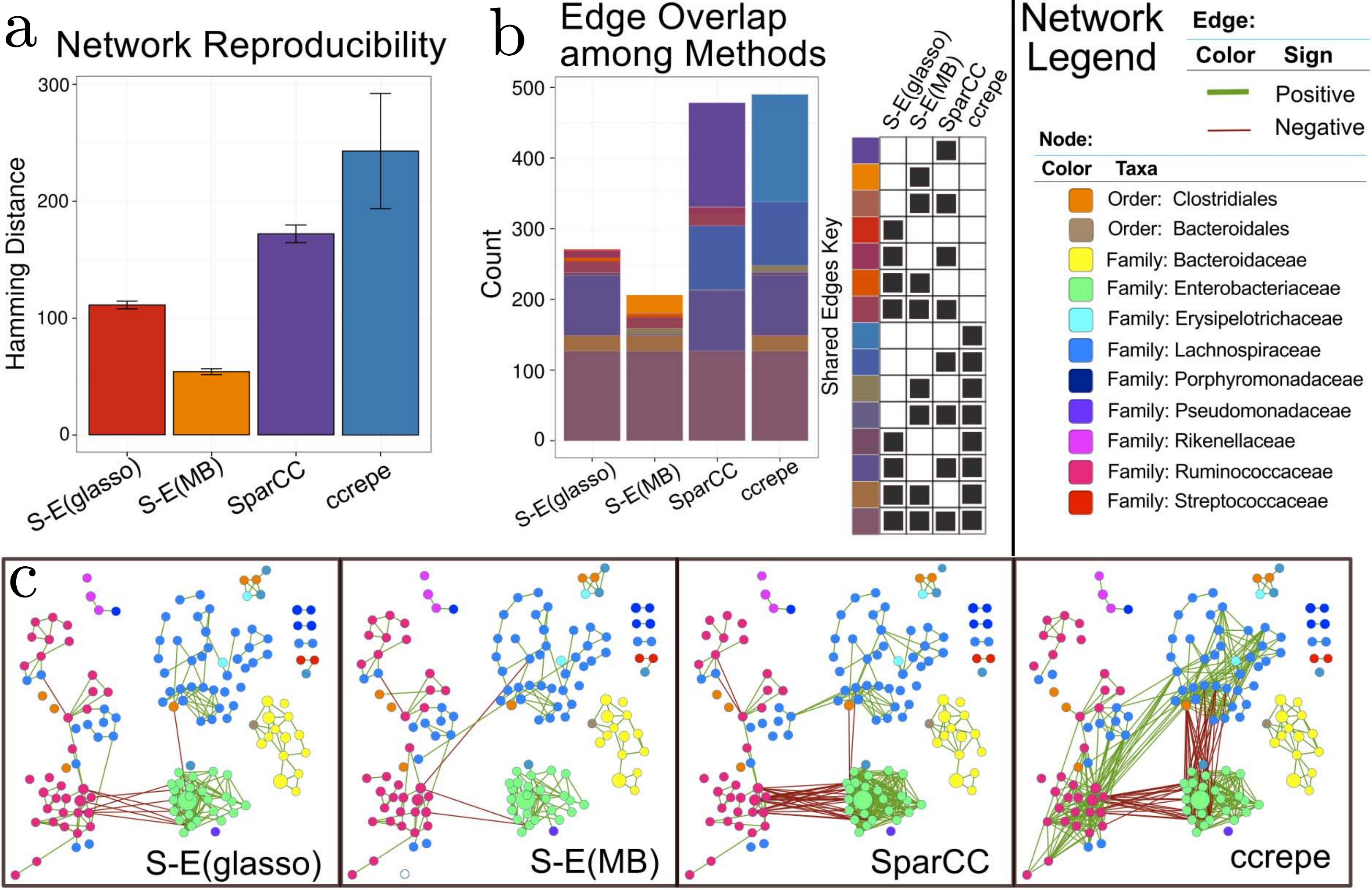}
\caption{\textbf{a}) Network reproducibility for inference methods (see main text for details). Bars represent mean Hamming distance, and errorbars are 95\% confidence intervals.  \textbf{b}) Visualization of edge overlap between networks inferred with SPIEC-EASI, SparCC, and CCREPE.  \textbf{c}) Network visualizations with OTU nodes colored by Family lineage (or Order, when the Family of the OTU is unknown), edges are colored by sign (positive: green, negative: red), and the node diameter proportional to the geometric mean of that OTU's relative abundance.}
\label{realnet}
\end{figure}

Figure \ref{realnet}a shows network reproducibility for SPIEC-EASI methods, SparCC, and CCREPE. The S-E(MB) has smallest the Hamming distance, followed by S-E(glasso), SparCC, and CCREPE. In S-E(MB), the edge disagreement is roughly 50 with very small error bars. At the other extreme, CCREPE edge disagreement is 250 edges and highly variable. 

These numerical experiments clearly demonstrate that SPIEC-EASI networks are more reproducible than other current methods. 
Finally, we use each inference method to construct a candidate American Gut microbiome association network from the unified dataset of size $n_1+n_2=558$ (Figure \ref{realnet}c). We analyze the differences between the reconstructed networks by quantifying the number of unique and shared predicted edges (Figure \ref{realnet}b). All four methods agree on a core network that consists of 127 edges. These edges are mostly found within OTUs of the same taxonomic group. This phenomenon, termed assortativity, has also been observed in other microbial network studies \cite{Faust:2012a}. Assortativity is one of the most salient features of the AGP-derived networks, and, for all networks, the assortativity coefficients for each network are close to unity (e.g., maximum assortativity, Figure \ref{assort}).  The SparCC network comprises about twice as many edges as the SPIEC-EASI networks. SparCC infers 147 distinct edges; these additional edges correspond to negative associations between OTUs of Ruminococcaceae (genus \emph{Faecalibacterium}) and Enterobacteriacae families (various genera) and a dense web of correlations within Enterobacteriacae OTUs.  Similarly, CCREPE identified 152 edges uniquely, with many negative edges between Enterobacteriaceae and Lachnospiraceae (genera: \emph{Blautia}, \emph{Roseburia} and unknown); additionally, CCREPE uniquely predicted positive edges between the Lachnospiraceae and Ruminococcaceae (genus: \emph{Faecalibacterium}). Both SPIEC-EASI methods produce relatively sparse networks by comparison. S-E(glasso) infers a total of 271 total edges (with one unique edge), and S-E(MB) infers 206 edges with 25 unique edges. In scale with edge predictions, both CCREPE and SparCC infer networks with large maximum degree (33 and 30, respectively), while the S-E(MB) and S-E(glasso) networks have a maximum degree of sixteen and eight, respectively (Figure \ref{assort}). However, even though CCREPE and SparCC predict a similar number of total edges, the global network properties are distinct. CCREPE predicts a higher maximum betweenness centrality and a larger number of nodes in the largest connected component (100).

In summary, analysis of the AGP networks suggests that the SPIEC-EASI inference schemes construct more reproducible taxon-taxon interactions than SparCC and CCREPE and infer considerably sparser model networks than the other two methods.  These observations may be explained as follows: SparCC and CCREPE aim to recover correlation networks, which contain both direct edges as well as indirect (e.g., spurious) edges (due to correlation alone).  SparCC and CCREPE may recover indirect edges less robustly than direct edges, an explanation that would be consistent with the Hamming distance reproducibility analysis.  In addition, all methods' resulting networks suggest that the topology of the American Gut association network cannot be attributed to a specific network class.  Instead, these networks are a mixture of band, scale-free, and cluster network type, and they exhibit high phylogenetic assortativity within highly connected components. 

\section*{Discussion}
\label{sec:discussion}
Inferring interactions among different microbial species within a community and understanding their influence on the environment is of central importance in ecology and medicine \cite{Faust:2012b,Longman:2013}. An ever increasing number of recent amplicon-based sequencing studies have uncovered strong correlations between microbial community composition and environment in diverse and highly relevant domains of life \cite{Gilbert:2010,Hmp:2012,deVos:2012,Gevers:2014,Lee:2014,Scher:2013}. These studies alone underscore the need to understand how the microbial communities adapt, develop, and interact with the environment \cite{Foster2008}. 
Elucidation of species interactions in microbial communities across different environments remains, however, a formidable challenge. Foremost, available high-throughput experimental data are compositional in nature, overdispersed, and usually underdetermined with respect to statistical inference. In addition, for most microbes few to no ecological interactions are known, thus the ecological interaction network must be constructed \textit{de novo}, in the absence of guiding assumptions and a set of "gold standard" interactions for validation.

To overcome both challenges, we present SPIEC-EASI (\textbf{S}parse \textbf{I}nvers\textbf{E} \textbf{C}ovariance Estimation for \textbf{E}cological \textbf{A}ssociation \textbf{I}nference), a computational framework that includes statistical methods for the inference of microbial ecological interactions from 16S rRNA gene sequencing datasets and a sophisticated synthetic microbiome data generator with controllable underlying species interaction topology. SPIEC-EASI's inference engine includes two well-known graphical model estimators, neighborhood selection \cite{Meinshausen2006} and sparse inverse covariance selection \cite{Yuan:2007,Banerjee2008,Friedman2008} that are extended by compositionally robust data transformations for application to the specific context of microbial abundance data.

The synthetic data pipeline was used to generate realistic-looking gut microbiome datasets for a controlled benchmark of SPIEC-EASI's inference performance relative to two state-of-the-art methods, SparCC \cite{Friedman2012} and CCREPE \cite{Gevers:2014}. We showed that neighborhood selection (S-E(MB)) outperforms SparCC and CCREPE in terms of recovery of taxon-taxon interactions and global network topology features under almost all tested benchmark scenarios, while covariance selection (S-E(glasso)) performs competitively with and sometimes better than SparCC and CCREPE.

Through our simulation study, we demonstrate that several other factors, in addition to total number of samples, affect network recovery. Foremost and in agreement with theoretical results from high-dimensional statistics \cite{Ravikumar:2010,Ravikumar2011,Tandon2014}, network topology has a significant impact, as network recovery performance is nearly doubled from scale-free to cluster to band (Figure \ref{PRboth}) for fixed sample size, number of taxa, and condition number. 
We also demonstrated dependence to strength of direct interactions (and thus strength of correlations) within a given network. 
Our simulation study provides the community with rough guidelines for requisite sample sizes, given state-of-the-art network inference and basic assumptions about the underlying network. This is of obvious importance to experimental design and the estimation of statistical power.
Here, we used the synthetic data pipeline to generate datasets characteristic of the gut microbiome. However, the SPIEC-EASI data generator is generic and therefore enables  researchers to generate synthetic datasets that resemble microbiome samples in terms of taxa dispersion and marginal distributions from their field of research, such as soil or sea water ecosystems \cite{Gilbert:2010}.

Our application study on real American Gut Project (AGP) data revealed that inference with SPIEC-EASI produced more consistent and sparser interaction networks than SparCC and CCREPE. In addition, our AGP network analysis revealed several biologically relevant observations. Specifically, we observed that OTUs were more likely to interact with phylogenetically related OTUs (Figure \ref{realnet}c and Figure \ref{assort}). In addition, our gut microbial interaction networks appear to be a composite of network types, as we find evidence for scale-free, band-like, and cluster subnetworks.  

An important advantage of neighborhood and covariance selection as underlying inference frameworks is their ability to include prior knowledge about the underlying data or network structure from independent scientific studies in a principled manner. For example, in the neighborhood selection scheme, the standard LASSO approach can be augmented by a group penalty \cite{Yuan:2006} that takes into account \textit{a priori} known group structure. The assortativity observed in our gut microbial interaction networks suggests that such a grouping of OTUs based on phylogenetic relationship might improve inference. Moreover, if verified species interactions are available for a certain microbial contexts, this knowledge can be included in covariance and neighborhood selection by relaxing the penalty term on these interactions. This strategy has already been fruitfully applied to inference of similarly high-dimensional transcriptional regulatory networks \cite{Greenfield:2013}. Finally, in agreement with theoretical and empirical work in high-dimensional statistics, our synthetic benchmark results confirmed that networks with scale-free structures elude accurate inference even if the underlying network is globally sparse. Recent modified neighborhood \cite{Tandon2014} and covariance selection \cite{Liu:2011} schemes improve recovery of scale-free networks and can be conveniently included into SPIEC-EASI. 


Finally, although the main focus of this work is inference of microbial interaction networks, estimation of the regularized inverse covariance matrix with S-E(glasso) will be key to addressing several other important questions arising from microbiome studies. For example, statistical methods to infer which taxa are responsive to design factors in 16S gene amplicon studies is an active area of research.  Most methods test each taxon independently one-at-a-time (see \cite{McMurdie2014} and references therein) even though taxa are actually highly correlated and thought to ecologically interact. Inference of taxa responses from 16S rRNA gene sequencing datasets could be improved by modeling this correlation structure through incorporation of the inverse covariance matrix into the statistical model \cite{Lin:2014}. 

Other, more complex questions are motivated by a desire to understand why and how ecosystems evolve with time.
In the dynamic modeling setting, association networks have already been successfully used as an underlying structure to fit a differential-equation-based model of gut microbiome development in mice \cite{Marino2014}. Thus, association networks provide the underlying topology for dynamic models, which can be used to develop hypotheses about how the ecosystems might respond to specific perturbations \cite{Foster2008}.

In conclusion, SPIEC-EASI is an improvement over state-of-the-art methods for inference of microbial ecological networks from microbiome composition datasets.  We demonstrate this through rigorous benchmarking with synthetic networks and also through application to a true biological dataset. In addition, the LASSO underpinnings of the SPIEC-EASI inference methods provide 
a flexible and principled mathematical framework to incorporate additional information about microbial ecological association networks as it becomes available, thereby improving prediction.
We anticipate that SPIEC-EASI network inference will serve as a backbone for more sophisticated modeling endeavors, engendering new hypotheses and predictions of relevance to environmental ecology and medicine.


\section*{Acknowledgments}
We would like thank Eric Alm and Jonathan Friedman for many helpful discussions and the manuscript reviewers for key improvements in the presentation of this work.


%
%
%

\newpage
\appendix
\section{Appendix}
%
%


\subsection{Method comparison}
\begin{table}[H]
\centering
{\small
\begin{tabular}{rllll}
  \hline
 & SPIEC EASI & SparCC & CCREPE & Pearson \\ 
  \hline
Underlying Metric & Conditional Independence & Correlation & Correlation & Correlation \\ 
  Compositional Correction & Yes & Yes & Yes & No \\ 
  Aitchison Measure & Inverse of clr covariance matrix & Aitchison Variation & - & - \\ 
  Network Assumptions & Network sparsity & Average correlation is zero & - & - \\ 
   \hline
\end{tabular}
}
\end{table}

Table to compare some of the features of SPIEC-EASI, SparCC, CCREPE and Pearson's correlation coefficient.
\label{methtable}

\subsection{Comparative marginal fits to American Gut Project count data}
\label{amgutfit}
We considered five common distributions for modeling OTU count data in SPIEC-EASI's synthetic data generator. As as target count data we used rounded common-scale normalized data from the American Gut Project (AGP). After filtering, a total of $p=205$ different OTU counts $w_i \in \mathbb{N}_{0}^{n}$ of sample size $n=558$ are available for marginal fitting. We fit the parameters of each distributional model  to each OTU count vector $w_i$ independently using maximum likelihood (ML) estimation. 

\subsubsection{Common count distributions}
We considered the following five common distributions to model univariate count data:

\paragraph*{Log-normal.}
The log-normal distribution is a continuous distribution of a random variable $U$ whose logarithm is normally distributed, $U \sim \ln \mathcal{N}(\mu, \sigma)$, where $\mu$ and $\sigma$ are the mean and standard deviation of $\ln(U)$, respectively. Its probability density function is given by

\begin{equation}
P(U = u) = \frac{1}{u\sigma \sqrt{2 \pi}}e^{-\frac{(\ln u - \mu)^2}{2\sigma^2}} \, ,
\end{equation}

To model discrete counts, the continuous values can be discretized by rounding to the nearest integer.

\paragraph*{Poisson.}
The Poisson distribution models a given number of discrete ``events"  occurring in a fixed interval (a  of an ecosystem). In ecology, a Poisson distributed random variable $U  \sim \ln \operatorname{Pois(\lambda)}$ describes the number of occurrences of a species across different ecosystem samples. The probability mass function is given by:
\begin{eqnarray}
P(U = u) =  \frac{\lambda ^u e^{-\lambda}}{u!}  \, ,
\end{eqnarray}
where $\lambda$ is the Poisson 'rate' parameter that determines both mean and variance.

\paragraph*{Zero-inflated Poisson.}
Zero-inflated models account for an excess of zero counts in real data that cannot be handled by a single component model. In ecology, count data may be skewed toward zero due to a preponderance of counts that fall below the sampling depth. The zero-inflated Poisson (ziP) model incorporates an additional term to account for this feature. The random variable reads:
\begin{eqnarray*}
U&\sim&\begin{cases}
0 & \mathrm{with\: prob.}\: \phi\\
\operatorname{Pois}(\lambda) & \mathrm{with\: prob.}\: 1-\phi
\end{cases}\\ \,
\end{eqnarray*}
and its mass function is given by:

\begin{eqnarray}
P(U = 0) &=& \phi + (1-\phi)e^{-\lambda}\\
P(U = u) & = & (1-\phi)\frac{\lambda ^u e^{-\lambda}}{u!}
\end{eqnarray}
where $\phi$ is the probability of obtaining an excess zero and $\lambda$ is the Poisson rate parameter as above. When $\phi = 0$, the ziP distribution reduces to the Poisson distribution.

\paragraph*{Negative Binomial.}
The negative binomial (NB) distribution arises as a hierarchical mixture of Poisson distributions that can model the variability from multiple sources (such as, e.g., biological replication and library preparation) \cite{McMurdie2014}. The NB distribution is thus more appropriate for handling overdispersion in the data.
We assume the means of the Poisson variables to be Gamma-distributed random variables with shape hyper-parameter $r$ and scale $\theta=p/(1-p)$. First, a random Poisson mean $\lambda$ is sampled from a Gamma distribution $\Gamma(r,\theta)$, and then a random variable $u$ from $\operatorname{Pois(\lambda)}$. The compact form of the mass distribution of a discrete NB random variable $U \sim \operatorname{NB(r; p)}$ then reads:
\begin{equation}
P(U = u) =  \frac{\Gamma(r+u)}{u!  \, \Gamma(r)} p^{u} (1-p)^{r} \, .
\end{equation}
Here, overdispersion is controlled by the parameter $r$. The NB model collapses to the Poisson model for $r \to \infty$. 

\paragraph*{Zero-inflated Negative Binomial}
Similar to the ziP model, the zero-inflated NB (ziNB) distribution can better account for excess zeros in the observed data. The standard NB distribution is augmented by a zero-inflation term. We denote a ziNB random variable by:
\begin{equation}
U\sim\begin{cases}
0 & \mathrm{with \: prob.} \, \phi \ \\
\operatorname{NB}(r; p) & \mathrm{with\: prob.}\: 1-\phi \, ,
\end{cases}\\
\end{equation}
where $\phi$ is the probability of obtaining an excess zero. The corresponding ziNB mass function reads:  
\begin{eqnarray}
P(U = 0) &=& \phi + (1-\phi) p^{u} \\
P(U = u) & = & (1-\phi) \frac{\Gamma(r+u)}{u!  \, \Gamma(r)} p^{u} (1-p)^{r} \, .
\end{eqnarray}
For $\phi = 0$ the ziNB model reduces to the NB distribution.

\subsubsection{Goodness-of-fit to the AGP data}
\label{sec:amgut_fit}
\begin{figure}[H]
	\centering
	\includegraphics[width=16cm]{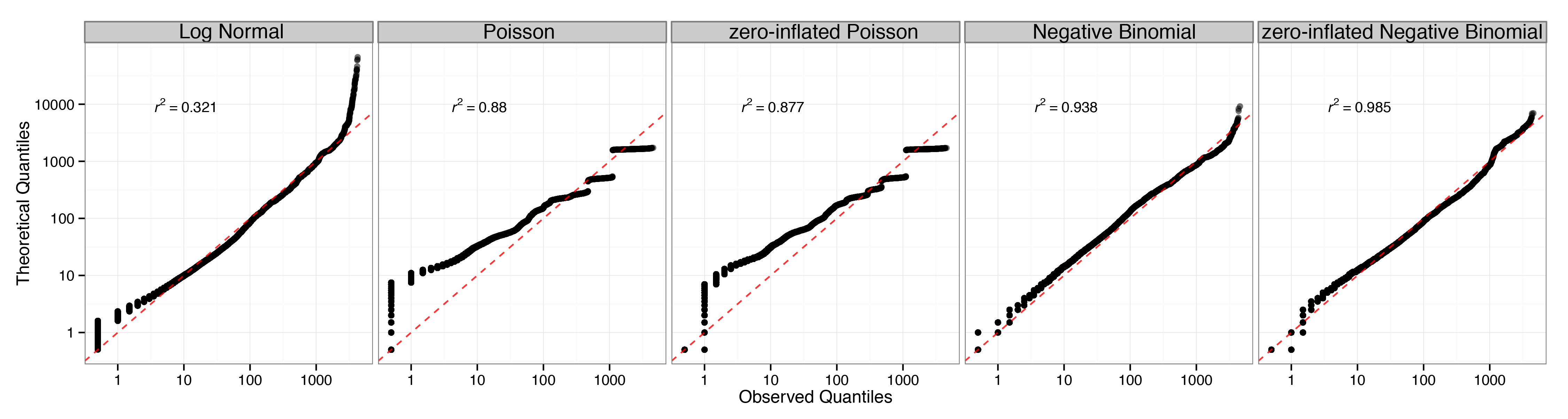}%
		\caption{QQ plots of AGP data fit to five different distributions (shown on a log-log scale for better visibility). The $r^2$ values correspond to the unscaled quantile-quantile relationships.}
		\label{amgutfit_fig}
\end{figure}

The AGP data was normalized using common-scale normalization, and fractional counts were rounded. We fit the count data to the five different statistical models and drew synthetic data (under null correlation between OTU marginals) from these distributions with the ML fitted parameters. 
Figure \ref{amgutfit_fig} shows QQ plots of synthetic and real data for the five different distributions.
The ziNB distribution ($r^2=0.985$) is the only model that can accurately model both tails of the OTU data. We thus use the ziNB as the standard setting in SPIEC-EASI's data generation model.


\subsection{Heatmaps of AGP and synthetic data}
\label{amgutfit_heatmap}
\label{amgutfit2_fig}
\begin{figure}[H]
\centering
\includegraphics[scale=.5]{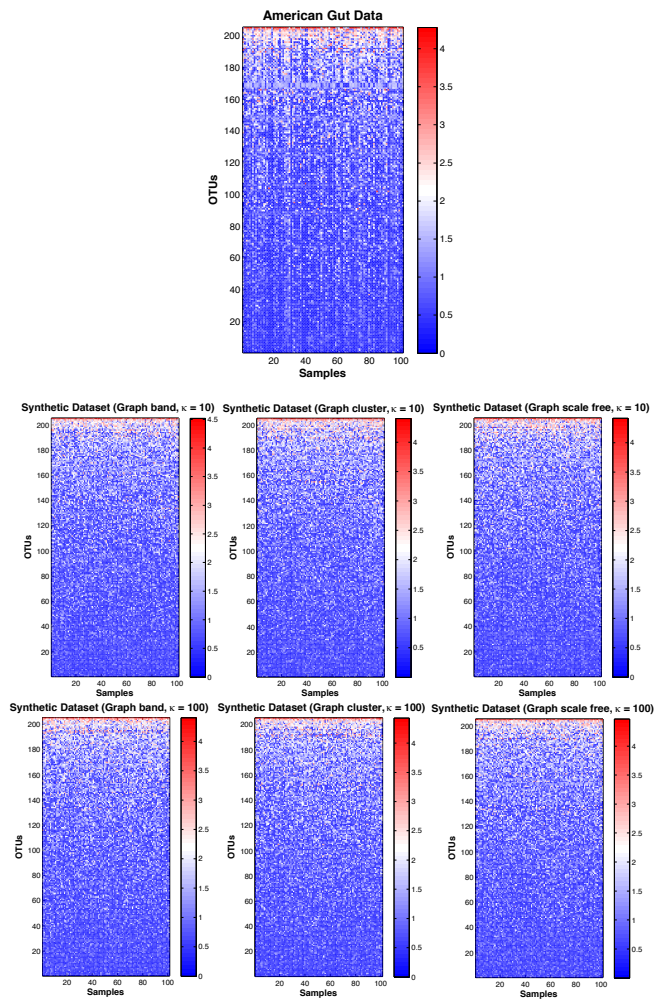}
\caption{Visual comparison of American Gut Project data with synthetically generated datasets.  Heatmaps show log10(counts).  For all network types (band, graph, scale-free), synthetic datasets are consistent with real datasets in terms of number of OTUs, number of samples, and OTU count abundances across samples.}
\end{figure}

\subsection{Relationship between target and empirical correlation after NorTA transformation}
\label{sec:norta_recov}
\label{norta_corr}
\begin{figure}[H]
\centering
\includegraphics[scale=.3]{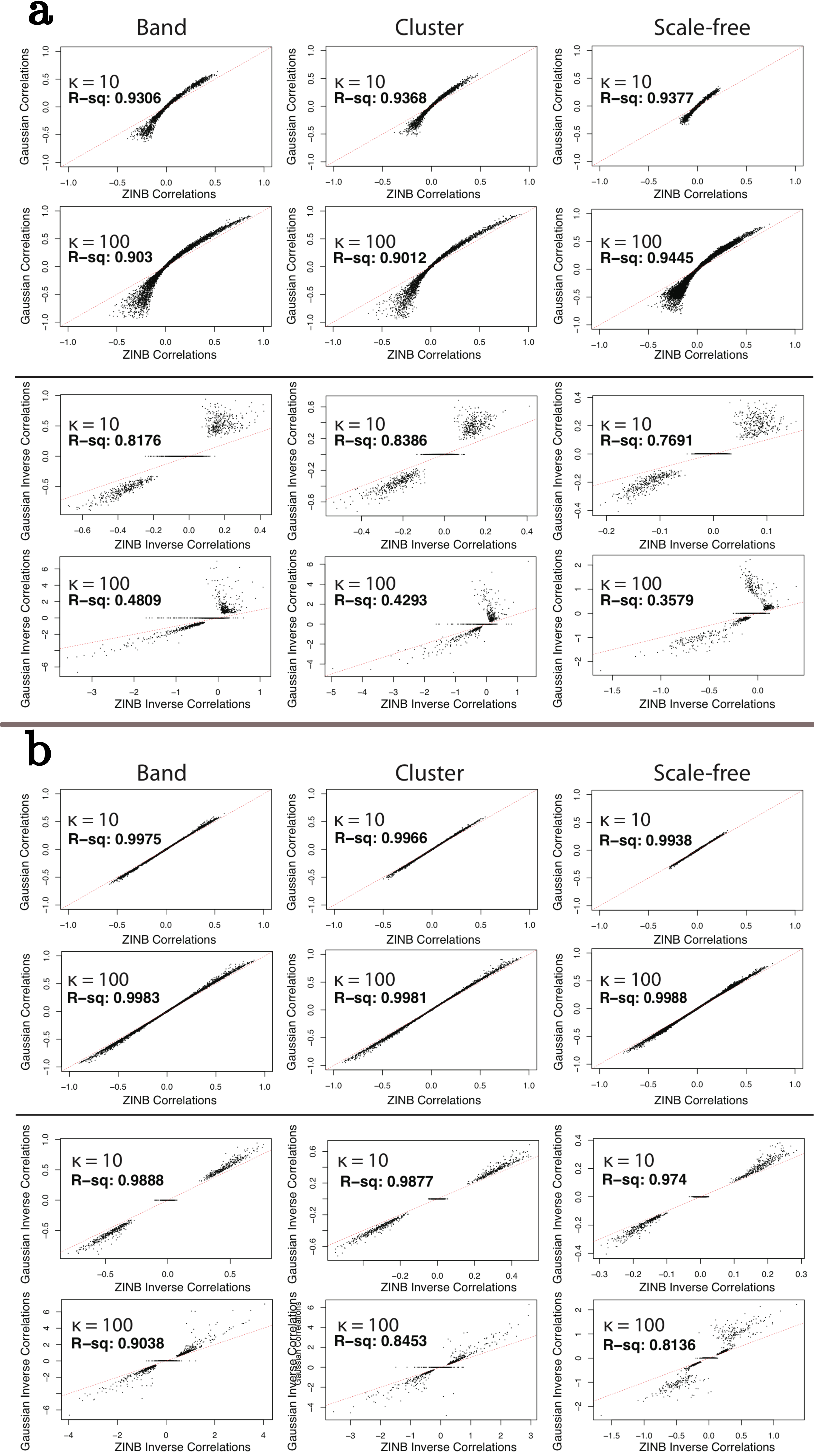}
\caption{The recovery of empirical Pearson correlations generated from the NORTA process, using zero-inflated Negative Binomial as a model (x-axis) verses the input multivariate Normal empirical correlations (upper panels) or inverse correlations (lower panels) on untransformed counts (\textbf{a}) or log-transformed counts (\textbf{b}). Simulated data are with $p=205$ OTUs, $n=20,000$ samples with $10$ replicates on each plot.}
\end{figure}

\subsection{Effect of Condition Number on Correlations Distribution}
\label{sec:cond_corr}
\label{cond_corr}
\begin{figure}[H]
\centering
\includegraphics[scale=.5]{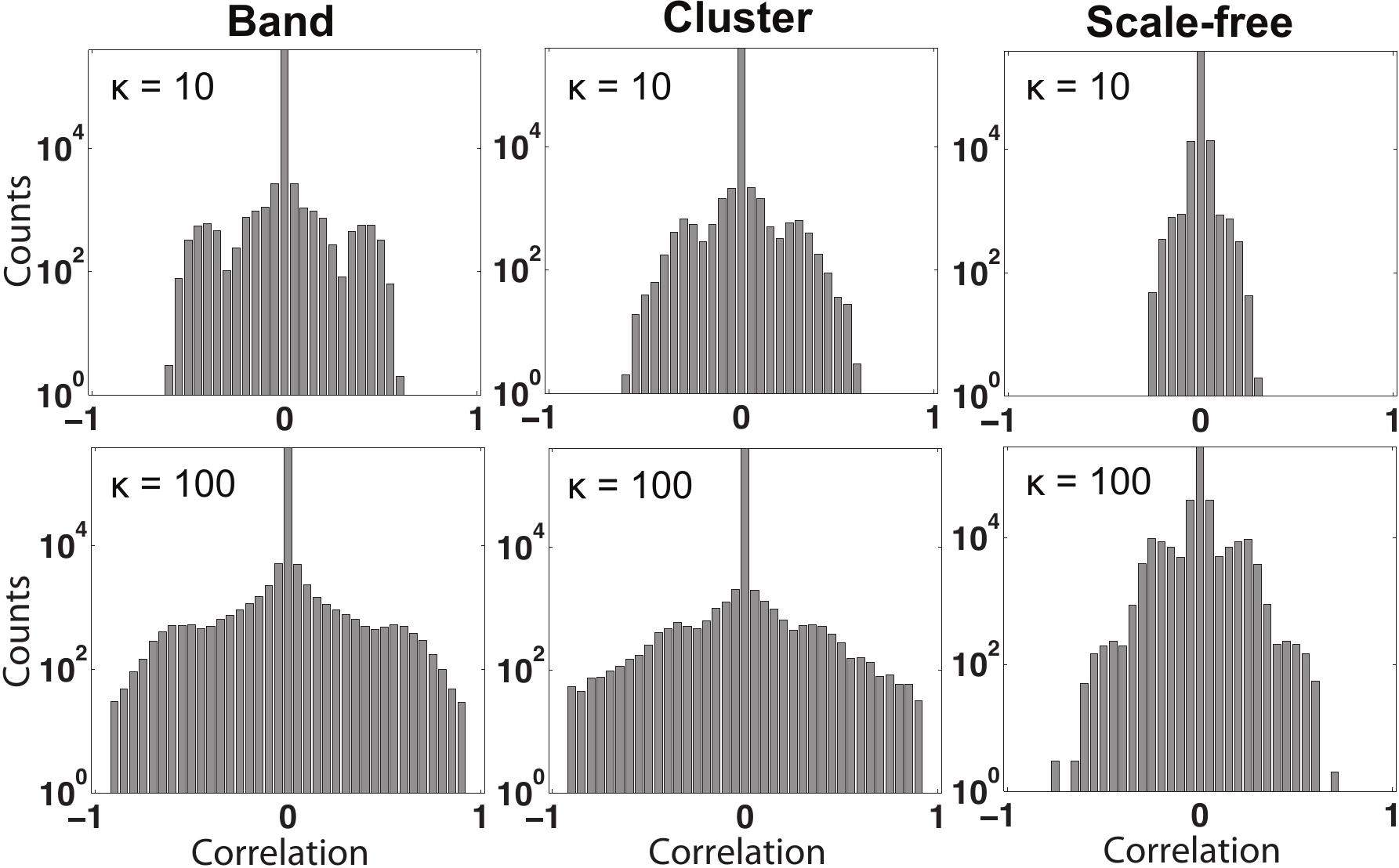}
\caption{The condition of a Precision matrix is the ratio of the largest to smallest eigenvalue/singular value. The relationship between condition number and correlation distribution for the synthetic networks; increasing condition number corresponds to increasing the strength of correlations in the network.}
\end{figure}

\subsection{Betweenness Centrality for Synthetic Networks}
\label{sec:betw_syn}
\label{hist_between}
\begin{figure}[H]
\centering
\includegraphics[scale=.6]{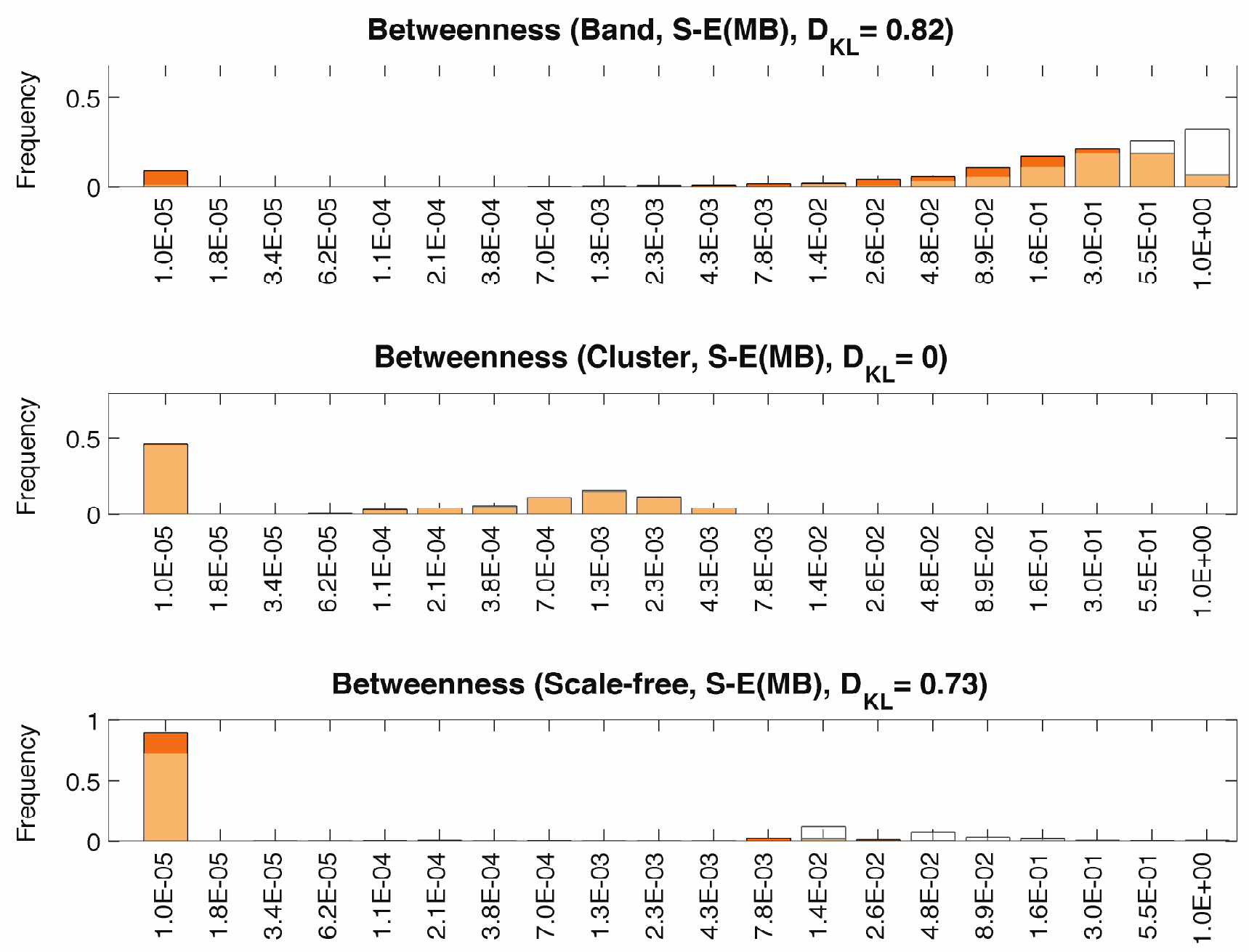}
\caption{Examples of betweenness centrality distributions for each network type (in white) overlayed with the distribution predicted by S-E(MB) (in orange) for $\kappa = 100$, $n = 1360$ samples, $p = 205$ OTUs.}
\end{figure}

\subsection{Recovery Performance of Betweenness Centrality}
\label{betweenness}
\begin{figure}[H]
\centering
\includegraphics[scale=.6]{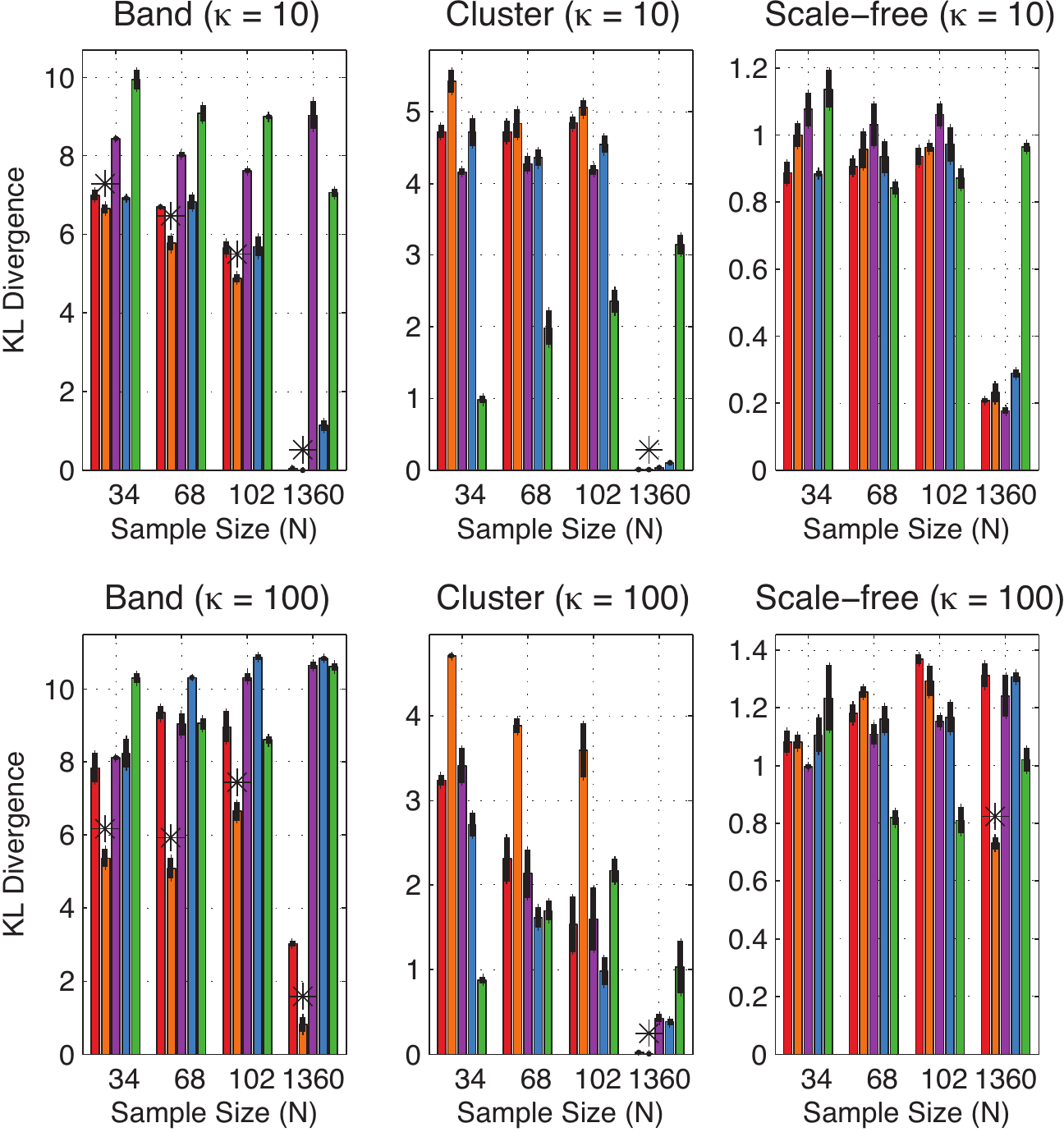}
\caption{Performance results for betweenness centrality (red = S-E(glasso) orange = S-E(MB), purple = SparCC, blue = CCREPE, green = Pearson).  Bars represent the average KL-divergence over three independent sets of synthetic datasets (7 datasets per set); error bars represent standard error. Asterisks indicate that an S-E method had siginificantly better recovery of the true betweenness centrality distributions ($p < 0.05$ for one-sided T tests in comparison to each control method).}
\end{figure}

\subsection{Geodesic Distance for Synthetic Networks}
\label{sec:geod_syn}
\label{geod_hist}
\begin{figure}[H]
\centering
\includegraphics[scale=.6]{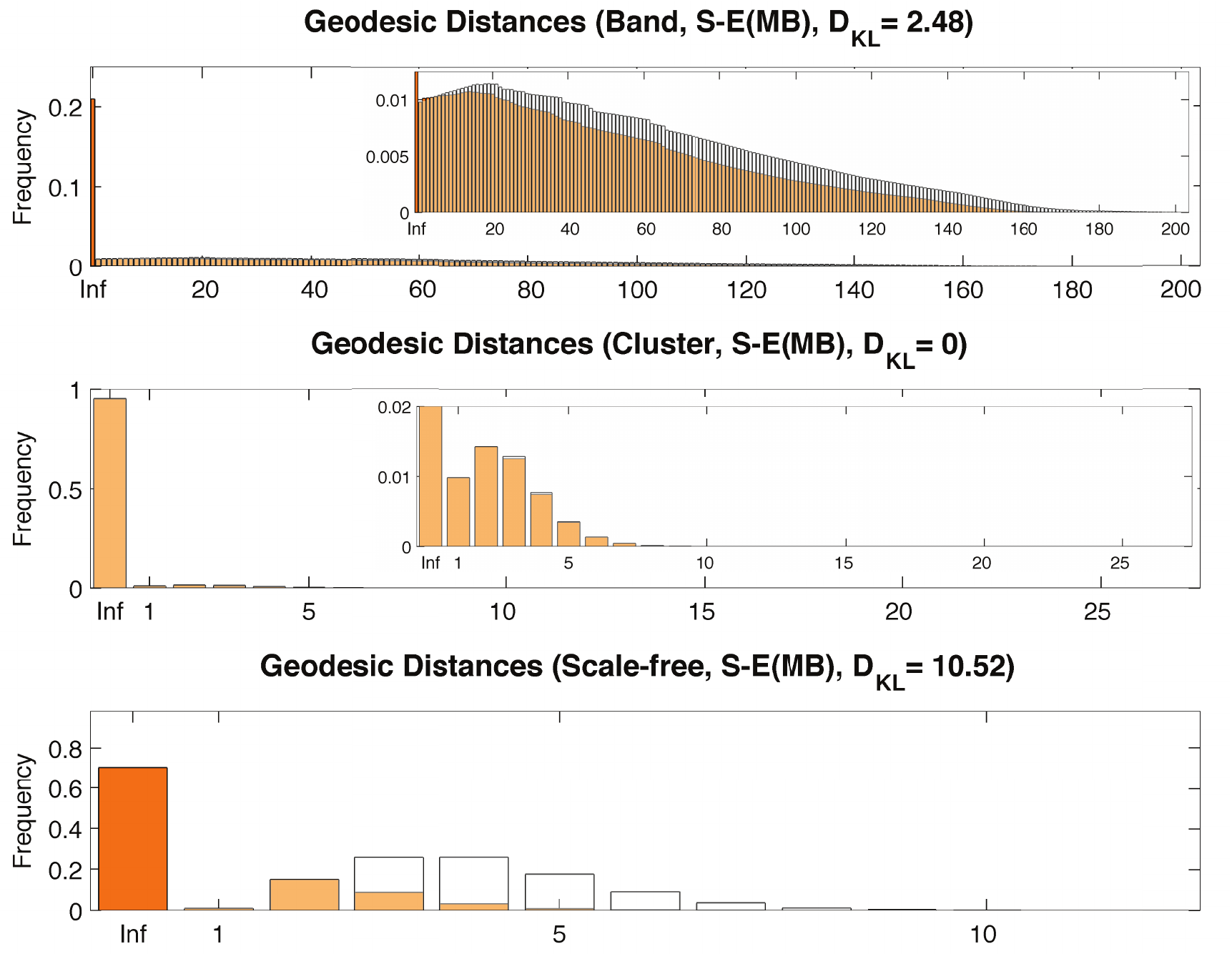}
\caption{Examples of geodesic distance distributions for each network type (in white) overlaid with the distribution predicted by S-E(MB) (in orange) for $\kappa = 100$, $n = 1360$ samples, $p = 205$ OTUs.}
\end{figure}

\subsection{Recovery Performance of Geodesic Distance}
\label{geodesic}
\begin{figure}[H]
\centering
\includegraphics[scale=.6]{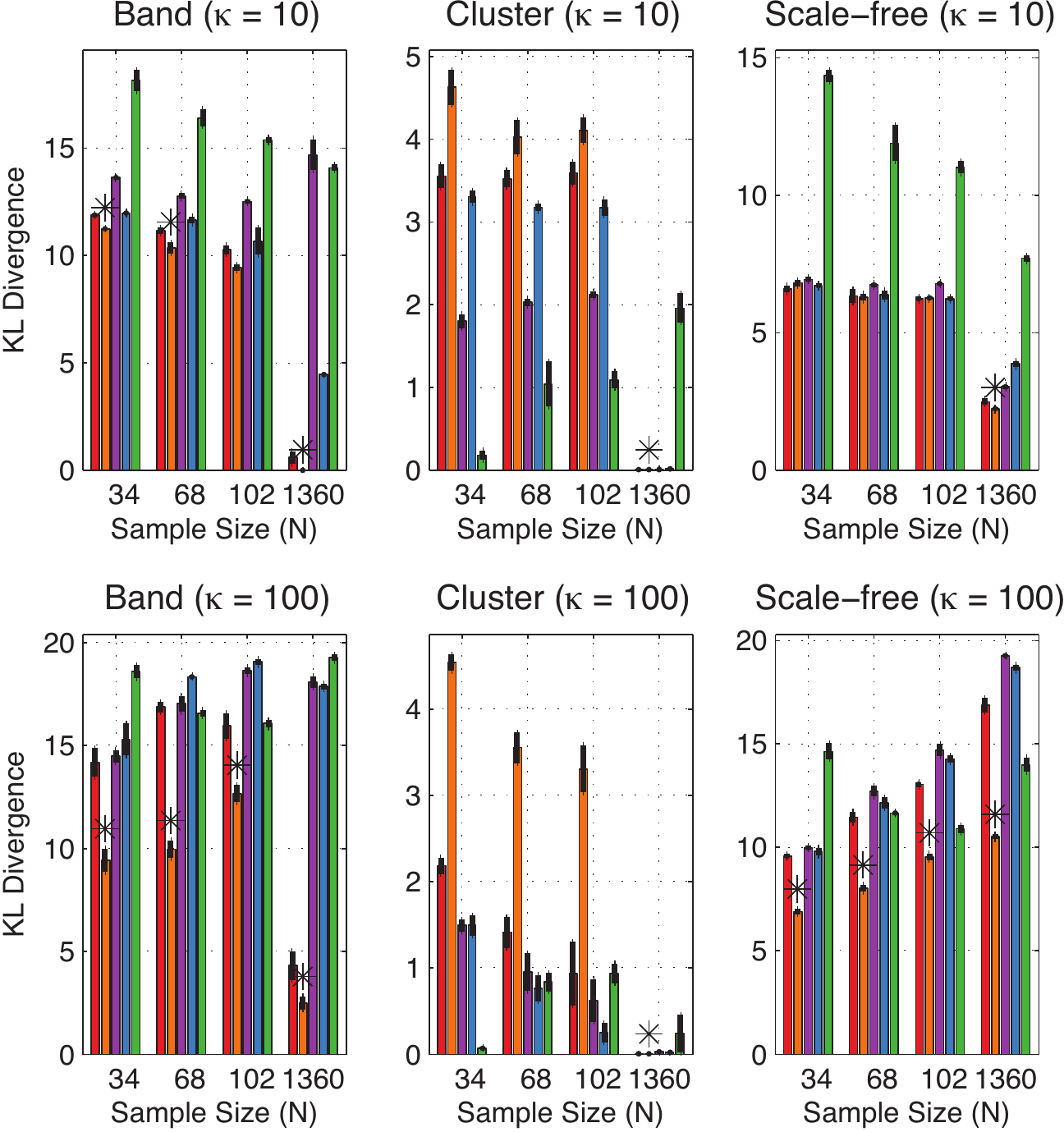}
\caption{Performance results for geodesic distance distributions (red = S-E(glasso), orange = S-E(MB), purple = SparCC, blue = CCREPE, green = Pearson).  Bars represent the average KL-divergence over three independent sets of synthetic datasets (7 datasets per set); error bars represent standard error. Asterisks indicate that an S-E method had significantly better recovery of the true geodesic distance distributions ($p < 0.05$ for one-sided T tests in comparison to each control method).}
\end{figure}

\subsection{Cluster Sizes for Synthetic Networks}
\label{clustsize_syn}
\label{clustersize}
\begin{figure}[H]
\centering
\includegraphics[scale=.6]{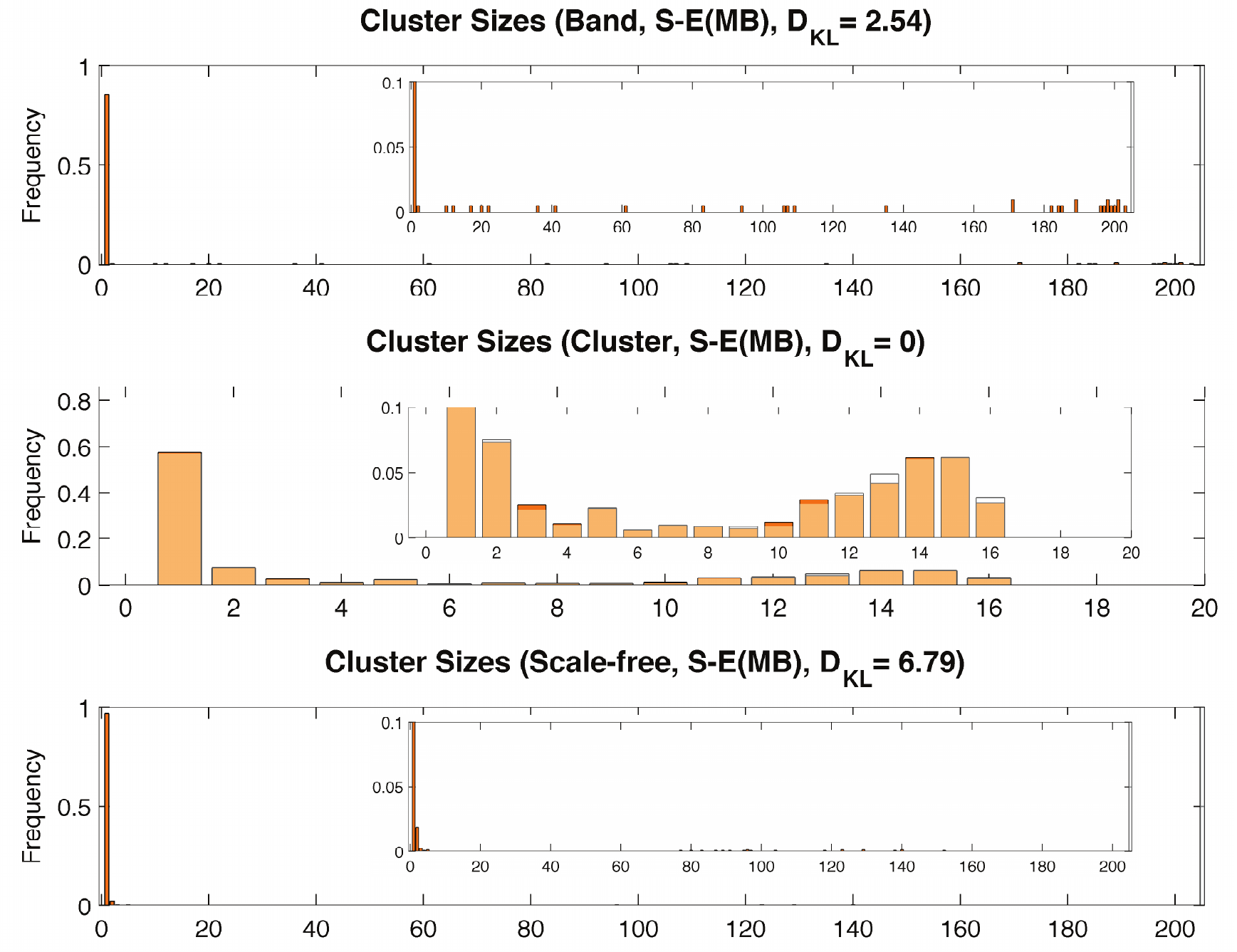}
\caption{Examples of cluster size distributions for each network type (in white) overlayed with the distribution predicted by S-E(MB) (in orange), for $\kappa = 100$, $n = 1360$ samples, $p = 205$ OTUs.}
\end{figure}

\subsection{Recovery Performance of Cluster Sizes}
\label{clustersize2}
\begin{figure}[H]
\centering
\includegraphics[scale=.6]{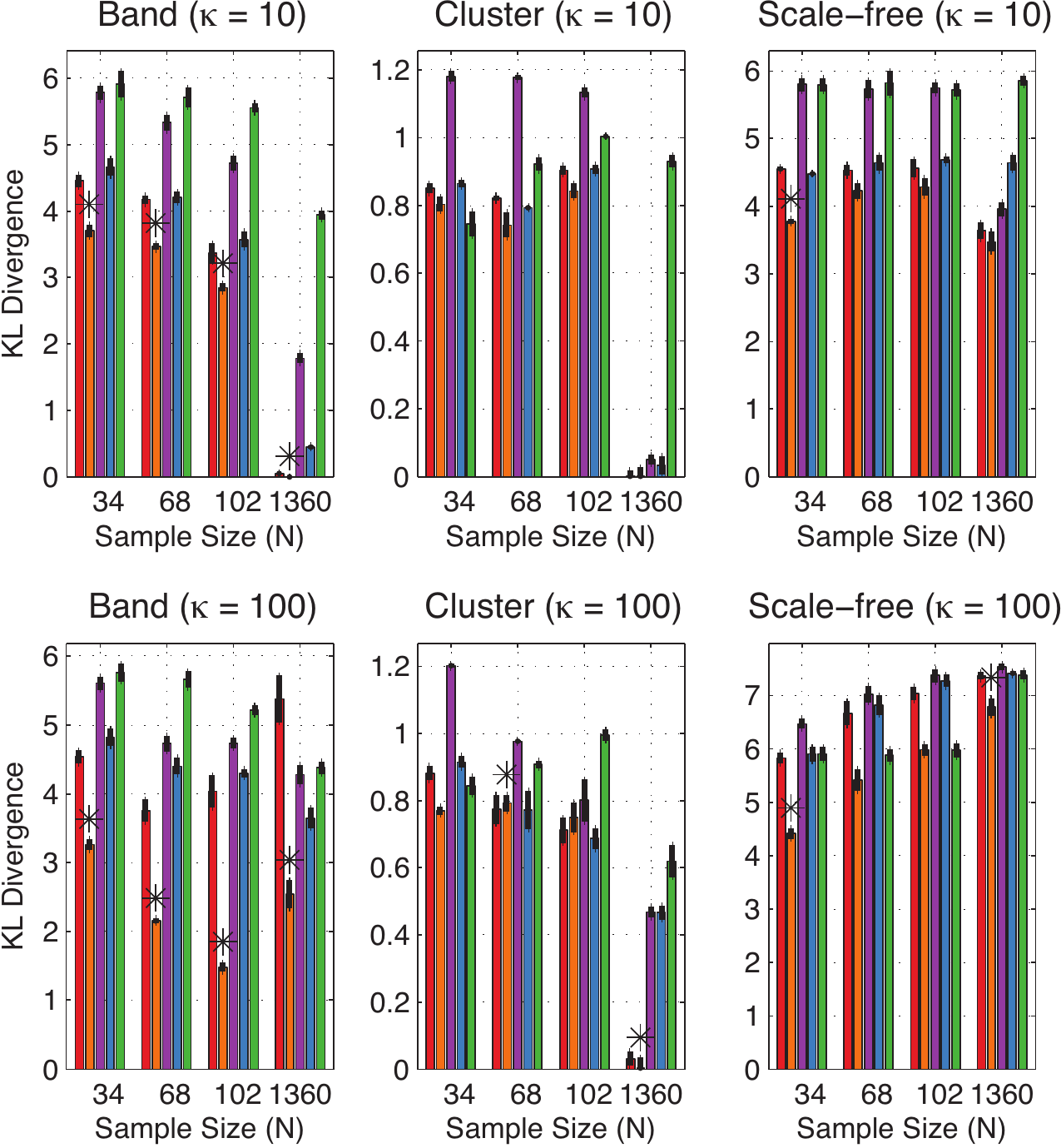}
\caption{Performance results for cluster size distributions (red = S-E(glasso), orange = S-E(MB), purple = SparCC, blue = CCREPE, green = Pearson).  Bars represent the average KL-divergence over three independent sets of synthetic datasets (7 datasets per set); error bars represent standard error. Asterisks indicate that an S-E method had significantly better recovery of the true cluster size distributions ($P < 0.05$ for one-sided T tests in comparison to each control method).}
\end{figure}

\subsection{Synthetic test case from SparCC}
\label{sparcc_benchmark}
\label{sparccfake}
\begin{figure}[H]
\centering
\includegraphics[scale=.5]{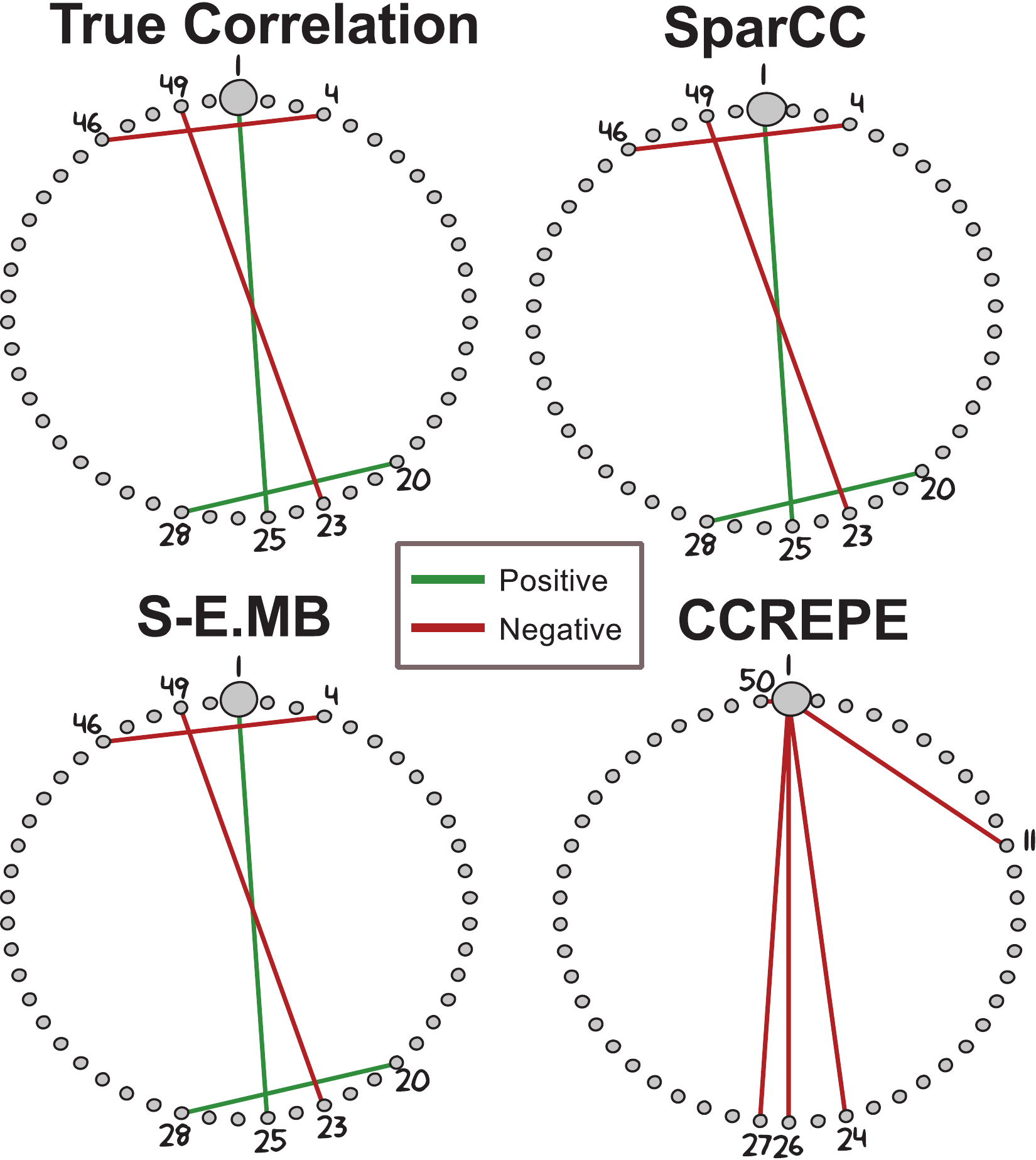}
\caption{Using the example data provided by the SparCC package, we inferred networks from SparCC (threshold correlation at $\pm.35$), S-E(MB) and CCREPE (thresholding q-value at $5*10^{-19}$). In this setting, SparCC and SPIEC-EASI correctly recover four true edges, including the association sign.}
\end{figure}

\subsection{Assortativity coefficients in inferred American Gut networks}
\label{assort}
\begin{figure}[H]
\centering
\includegraphics[scale=.4]{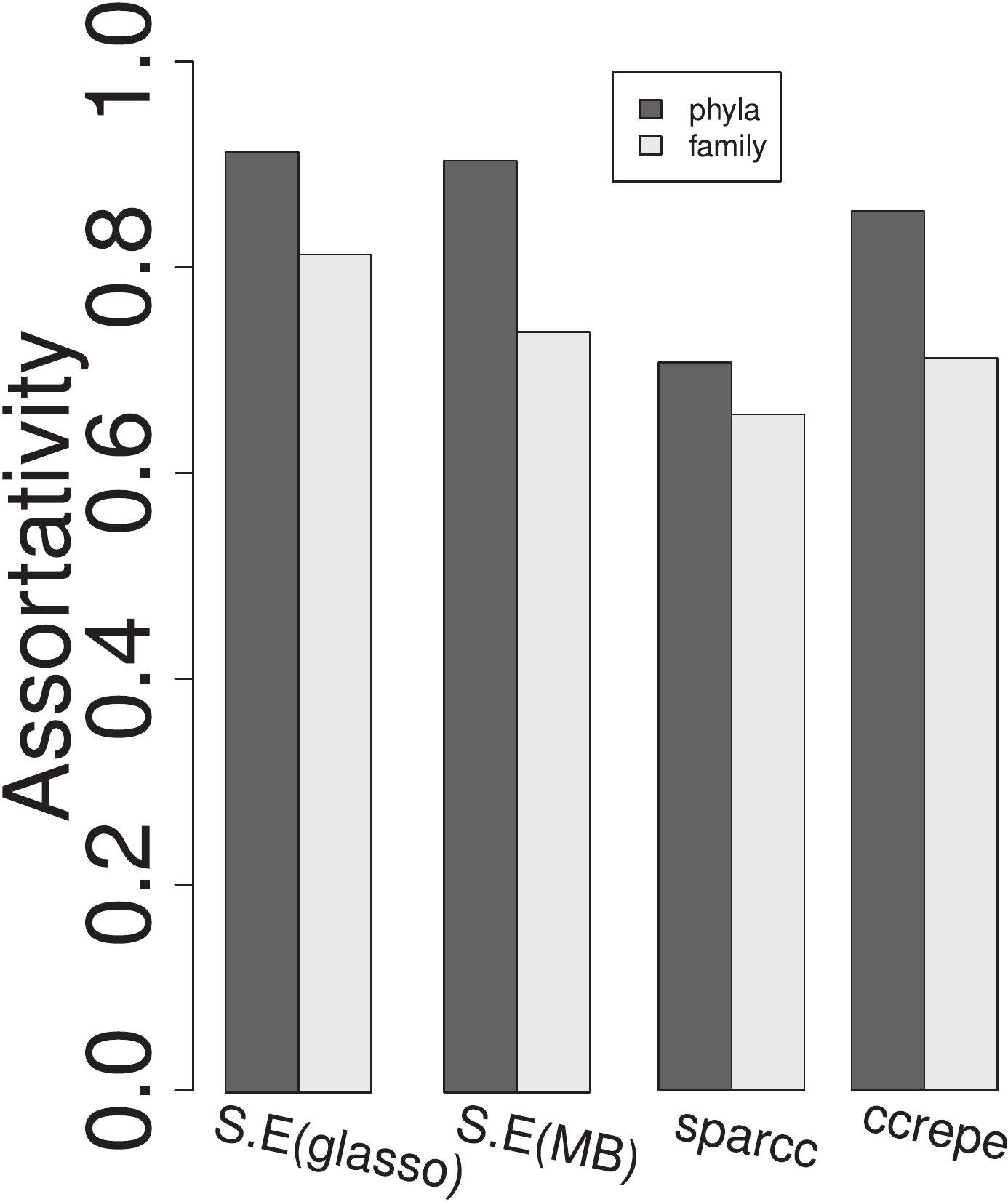}
\caption{Network assortativity coefficients at the Phyla and Family level for each of the four inference methods. Assortativity is a measure of the tendency for nodes to be connected with nodes of the same taxonomic class.}
\end{figure}

\end{document}